\documentclass[a4paper,12pt]{article}
\usepackage{jheppub}
\usepackage{orcidlink}
\usepackage{hyperref}
\usepackage[T1]{fontenc}
\title{\boldmath Thermodynamic properties of Quantum-Corrected AdS Black Hole with Phantom Global Monopoles}

\author[a]{B. Hamil\orcidlink{0000-0002-7043-6104},}
\author[b]{B. C. L\"{u}tf\"{u}o\u{g}lu\orcidlink{0000-0001-6467-5005},}
\author[c]{F. Ahmed\orcidlink{0000-0003-2196-9622}
,}
\author[d]{and Z. Yousaf\orcidlink{0000-0001-8227-2621}
}

\affiliation[a]{Laboratoire de Physique Mathématique et Subatomique, Faculté des Sciences Exactes, Université Constantine 1, Constantine, Algeria}
\affiliation[b]{Department of Physics, Faculty of Science, University of Hradec Kralove, Rokitanskeho 62/26, Hradec Kralove, 500 03, Czech Republic}
\affiliation[c]{Department of Physics, University of Science \& Technology Meghalaya, Ri-Bhoi, Meghalaya, 793101, India}
\affiliation[d]{Department of Mathematics, University of the Punjab, Quaid-i-Azam Campus, Lahore 54590, Pakistan}

\emailAdd{hamilbilel@gmail.com}
\emailAdd{bekir.lutfuoglu@uhk.cz}
\emailAdd{faizuddinahmed15@gmail.com}
\emailAdd{zeeshan.math@pu.edu.pk}

\abstract{In this paper, we introduce a metric ansatz designed to describe spherically symmetric quantum-corrected black hole (BH) space-time within an AdS space background, incorporating both an ordinary and a phantom global monopole. Our study focus into the thermodynamic properties of this BH, where we compute key parameters such as the Hawking temperature and specific heat capacity. We then proceed to analyze the effective potential of the system, considering both null and time-like geodesics, and investigate the shadow radius of the BH. Additionally, we calculate the emission rate of particles from the BH, providing insights into the energy dynamics. The geodesic equations of motion are explored to visualize the trajectories of massive particles within the BH. Throughout our investigation, we thoroughly examine how the inclusion of both ordinary and phantom global monopoles, combined with the quantum-corrected parameter, influences various thermal properties, the effective potential of the system, the BH shadow radius, energy emission rate, and the trajectories of massive particles.  Importantly, by generating figures that depict these phenomena, we emphasize the differences in results obtained with ordinary global monopoles and phantom ones, considering a range of quantum-corrected parameter values and small energy scale parameters, which allows us to discern the distinct effects of each type of monopole in the black hole's behavior.}


\arxivnumber{2404.16674}

\begin{document}

\maketitle

\small 

\section{Introduction}\label{sec:intro}

General Relativity (GR) is widely regarded as the most successful theory for describing gravitational interactions \cite{aa1, aa2}. In this context, finding exact solutions to the Einstein field equations (EFE) is crucial for addressing persistent fundamental issues, such as the existence of spacetime singularities. Although GR is robust from both theoretical and experimental standpoints,  there are several compelling reasons to explore alternative theories of gravity. 

From a theoretical perspective, two significant challenges remain in GR: (i) the presence of singularities \cite{aa3,aa4} and (ii) the inability to renormalize GR using standard quantization techniques \cite{aa5}. From an observational standpoint, the discovery of the Universe's dark sectors highlights the necessity for new fundamental physics. This underscores the need to connect the ultraviolet (UV) and infrared (IR) sectors, potentially through modifications of GR that incorporate quantum features. To gain insights into possible modifications of GR, leveraging exact solutions of the EFE is essential. Some of the most important solutions to the EFE, despite being well-understood and extensively studied, still present challenges. Addressing these challenges may involve considering alternative theories of gravity, such as (i) asymptotically safe gravity applied to BHs \cite{aa6,aa7,aa8}, (ii) improved BH formalisms \cite{aa10, aa11, aa12}, and (iii) scale-dependent gravity \cite{aa13, aa14, aa15, aa16, aa17, aa18, aa19, aa20, aa21, aa22, aa23, aa24, aa25, aa26, aa27, aa28, aa29, aa30, aa31, aa32, aa33}. Other theories include Lovelock gravity, Brane world cosmology, $f(\mathcal{R})$-gravity, $f(\mathcal{T})$-gravity, $f(\mathcal{G})$-gravity,  $(\mathcal{R}, \mathcal{T})$-gravity, and $f(\mathcal{R}, \mathcal{G})$-gravity, among others (see, for example \cite{nn5, nn6, nn7, nn8, nn9,  nn11} and related references therein).

The investigation of gravitational waves (GWs) has played a pivotal role in cosmology and astrophysics since the formulation of the theory of general relativity in the 20th century. The Laser Interferometer Gravitational-Wave Observatory (LIGO) achieved a historic milestone by successfully detecting GW signals originating from binary BH mergers \cite{nn1, nn2, nn3, nn4}. Subsequently, the LIGO/Virgo collaborations extended this achievement to encompass mergers involving BH and neutron stars, ushering in a new era of multi-messenger astronomy. Notably, the emergence of GW signals from such events has spurred heightened interest in primordial BH (PBH) as they are posited to explain binary BH mergers. It is worth mentioning that BH stands as a crucial entity in the realm of quantum gravity. Yet, the direct detection of four-dimensional BH within particle accelerators remains an elusive pursuit. The staggering energy requirement, on the order of the Planck energy ($\sim 10^{19}$ GeV), exceeds current technological capabilities by a wide margin, dampening hopes of immediate observation. However, the prospect brightens in the presence of large extra dimensions. In such a scenario, the effective Planck scale can be lowered to the TeV range, rendering experiments feasible shortly. Type I and Type II string theories are characterized by this reduction in the Planck scale, which is brought about by localizing standard model particles in the realms of the D-brane while allowing gravity to propagate freely in the higher-dimensional bulk \cite{pp1}. Some recent investigations on BH physics can be found in \cite{mm1,mm2,mm3,mm4,mm5}. 

The concept of obtaining a consistent quantum theory of gravity through BH thermodynamics has been proposed. Viewing the geometric aspects of BH as thermodynamic variables offers a potent framework for constructing such a theory. Recognizing BH as a thermodynamic entities has profoundly influenced our comprehension of gravitational theory and its interplay with quantum field theory. Recent advancements in gauge/gravity duality highlight the importance of thermodynamics for BH \cite{pp2,pp3,pp4,pp5,pp6,pp7,pp8,pp9,pp10,pp11,pp12}. Moreover, seminal works by Hawking and Page, which explored phase transitions of asymptotically AdS BH \cite{SWH}, along with Witten's contributions on similar subjects \cite{pp13}, have further underscored the importance of BH thermodynamics. Various approaches, grounded in different ensembles, can be employed to study BH thermodynamics. For example, studying thermal stability within the canonical ensemble provides information about the sign of heat capacity, which controls whether BH is thermally stable or unstable. The points of the phase transition and bound states are represented by the divergences of heat capacity and roots, respectively. Consequently, BH thermodynamics and their thermal stability have been extensively explored in the literature \cite{pp14, pp15, pp16, pp17, pp18}. Pioneered by Bekenstein {\it et al.} \cite{ss1, ss2}, the widely accepted theory states that BH propagates as black bodies, with entropy being related to the horizon’s area. The BH, even ones considerably larger than the Planck scale, are now generally accepted to have entropy proportionate to their horizon area \cite{ss1, ss2, ss3, ss4, ss5, ss6}. This prompts an intriguing question: When BH shrinks in size, what becomes the leading-order corrections?

Various attempts have been made to address this question. For example, utilizing a modified version of the asymptotic Cardy formula for BTZ, string-theoretic, and other BH, whose microscopic degrees of freedom are described by an underlying Conformal Field Theory (CFT) \cite{ss7}, main modifications with logarithmic behavior have been revealed. Furthermore, taking matter fields into account in BH backgrounds results in logarithmic improvements to the entropy of BH at the leading order \cite{ss8, ss9, ss10}. Similarly, the leading-order correction to BH entropy is found to be logarithmic when considering the string-BH correspondence \cite{ss11, ss12} and utilizing the Rademacher expansion of the partition function \cite{ss13}.

The study of quantum-corrected BHs is crucial for addressing some of the most fundamental problems at the intersection of GR and quantum mechanics \cite{Kazakov1994}. Classical GR predicts the formation of singularities, where the spacetime curvature becomes infinite and the laws of physics break down. These singularities, most notably within black holes, signal the need for a quantum theory of gravity that can resolve such inconsistencies \cite{Kim2012}. Moreover, BHs serve as an ideal laboratory for testing the effects of quantum gravity due to their extreme curvature and strong gravitational fields \cite{Husain}. By incorporating quantum corrections into BH solutions, we gain valuable insights into phenomena such as Hawking radiation, the information paradox, and the potential resolution of singularities \cite{Shahjalal2019, Bezerra2019, Konoplya2020}. These corrections are essential for understanding spacetime's quantum structure and exploring possible bridges between GR and quantum field theory \cite{Badawi}. As a result, studying quantum-corrected BHs may provide a path toward a more complete theory of gravity that unifies the ultraviolet and infrared regimes, ultimately offering deeper insights into both cosmology and high-energy astrophysical processes \cite{Hhao}. Investigations into the leading-order corrections to BHs \cite{Barman}, and their thermodynamics are currently of significant interest \cite{Fatima2024, Sadeghi, Frago}. Recent studies have explored the effects of quantum corrections on the thermodynamics and stability of various BH solutions, as reported in \cite{ss14,ss15,ss16, BH1, Hao, BHBCL1}. The corrected thermodynamics of dilatonic BH  solutions have also been discussed, revealing a universal form of the correction term \cite{ss17}. Furthermore, investigations into the corrected thermodynamics of BH from the perspective of partition functions have been conducted \cite{ss13}. Quantum gravity effects on the thermodynamics and stability of Ho\u{r}ava-Lifshitz BH have been analyzed \cite{ss18}, as well as investigations into modified Hayward BH, revealing correction terms that reduce pressure and internal energy \cite{ss19}. { Some recent studies of black holes in general relativity and modified gravity theories with and without quintessence field and other configuration were reported in Refs. \cite{GM1,GM2,GM3,GM4,GM5,GM6,GM7,GM8,GM9,GM10,GM11}.}

The line-element describing a spherically symmetric BH with global monopoles (ordinary and phantom) in the spherical coordinates $(t, r, \theta, \phi)$ is given by \cite{SCJJ, AHEP}
\begin{equation}
ds^2=-\mathcal{B}(r)\,dt^2+\frac{dr^2}{\mathcal{B}(r)}+r^2\,(d\theta^2+\sin^2 \theta\,d\phi^2), \label{a3}
\end{equation}
with the function
\begin{equation}
    \mathcal{B}(r)=\Big(1-8\,\pi\,\eta^2\,\xi-\frac{2\,M}{r}\Big),\label{a4}
\end{equation}
where $\eta, M$ and $\xi$ stand for the energy scale of symmetry breaking, the quantity of matter, and BH kinetic energy, respectively. The scenario mediated by $\xi=1$ describes an arena of a regular global monopole that arises from the scalar field’s non-negative and non-zero kinetic energy \cite{MB}. However, the phantom global monopole is generated by selecting $-1$ value of $\xi$, thereby relating it with the scalar field’s negative kinetic energy.

On considering the quantum fluctuations associated with the spherically symmetric manifold, the corrections of 2D dilaton gravity theory can be observed mediated from the Einstein-Hilbert action accompanied by the 4D interaction theory \cite{DIK}. Under such circumstances, it is feasible to renormalize a gravitational theory such as the 2D dilaton gravity, which was proposed by Kazakov and Solodukhin. In Ref. \cite{wu}, authors presented a quantum-corrected Schwarzschild BH with an AdS background and investigated the thermodynamic properties. The line element describing this AdS background BH is given by  
\begin{equation}
ds^2=-f(r)\,dt^2+f^{-1}(r)\,dr^2+r^2\,(d\theta^2+\sin^2 \theta\,d\phi^2), \label{a5}
\end{equation}
where 
\begin{equation}
    f(r)=\frac{1}{r}\,\sqrt{r^2-\alpha^2}-\frac{2\,M}{r}+\frac{(r^2-\alpha^2)^{3/2}}{r\,\ell^2},\label{a6}
\end{equation}
where $\Lambda=-\frac{3}{\ell^2}$ is the cosmological constant and $\alpha=4\,\ell_{p}$ is a small correction that describes the behavior of spherical symmetric quantum fluctuations.   

\section{Quantum-corrected AdS BH with phantom global monopoles}\label{sec:1}

We aim to investigate a spherically symmetric AdS background BH with phantom global monopoles (ordinary and phantom one) taking into account the quantum corrections. Therefore, we begin this section by introducing the line-element ansatz of this spherically symmetric BH in the coordinates $(t, r, \theta, \phi)$ with ordinary and phantom global monopole taking into account the quantum correction given by
\begin{eqnarray}
ds^2=-\mathcal{F}(r)\,dt^2+\frac{dr^2}{\mathcal{F}(r)}+r^2\,(d\theta^2+\sin^2 \theta\,d\phi^2),\label{b}
\end{eqnarray}
where $\mathcal{F}$ is given by
\begin{equation}
    \mathcal{F} (r)=\frac{1}{r}\,\sqrt{r^2-\alpha^2}-\frac{2\,M}{r}-8\,\pi\,\eta^2\,\xi+\frac{(r^2-\alpha^2)^{3/2}}{r\,\ell^2}\, .\label{b1}
\end{equation}

One can see that for zero quantum-correction parameter $\alpha = 0$,  $\ell \to \infty$, and $\xi=1$, we will find a space-time describing global monopole geometry which was obtained by Barriola {\it et al.} \cite{MB}. Moreover, for $\alpha = 0$ and $\xi = 0$, one will find a spherically symmetric AdS background BH or a space-time of Schwarzschild-AdS BH \cite{SWH}. Furthermore, for $\xi=0$, the quantum-corrected AdS Schwarzschild BH geometry described by the space-time (\ref{a5}) is recovered \cite{wu}. In this work, we take a step further by examining the effects of both ordinary and phantom global monopoles, as well as other parameters involved in the geometry (\ref{b}), on the thermodynamic properties and geodesic motions of test particles. We analyze the outcomes of these investigations to gain a deeper understanding of the underlying physical phenomena.

In this paper, our objective is to conduct a detailed investigation of spherically symmetric AdS background quantum-corrected BH featuring both ordinary and phantom global monopoles. We structure our investigation as follows: Firstly, we study the thermodynamic properties of the quantum-corrected BH and analyze the physical parameters associated with the system. Next, we analyze how the effective potential of the system and the shadows of BH are influenced by the various parameters involved. Afterward, we calculate the energy emission rate of the chosen BH and illustrate graphically how ordinary and phantom global monopoles alter their value. Finally, we focus on the geodesic motions and discuss the trajectories of massive and light-like particles within this framework. 

\subsection{Thermodynamic analysis}\label{subsec:1}

At first, we attempt to obtain an analytic form of the event horizons by determining the roots of the lapse function. { The horizon condition at $r=r_{+}$ is given by \cite{GM2,GM10,GM11}}
\begin{equation}
\mathcal{F}(r)\Big\vert_{r=r_{+}}=0.
\end{equation}
Using the lapse function (\ref{b1}), we find the following polynomial equation for $r_{+}$ given by 
\begin{equation}
\frac{1}{r_{+}}\,\sqrt{r_{+}^{2}-\alpha^{2}}-\frac{2\,M}{r_{+}}-8\,\pi \,\eta ^{2}\,\xi +\frac{8\,\pi\, P\, (r_{+}^{2}-\alpha^{2})^{3/2}}{3\,r_{+}}=0 \label{d2}
\end{equation}
that does not provide an analytic solution for $r_{+}$, and numerical methods must be employed, where $P=\frac{3}{8\pi \ell ^{2}}$ is the thermodynamic pressure. It is better to mention here that for $\xi=0$ in eq. (\ref{d2}), one can find a cubic equation for $x=r^{2}_{+}$, that is, $c_{1}\,x^3+c_2\,x^2+c_3\,x+c_0=0$ which can easily be solved, where $c_i\,(i=0,1,2,3)$ are constants. Thus, we can affirm from Eq. (\ref{d2}) that $r_{+}$ is related with parameters $P, M, \alpha$, and $\xi$. 

\begin{figure}[htb!]
\centering
\includegraphics[width=0.35\textwidth]{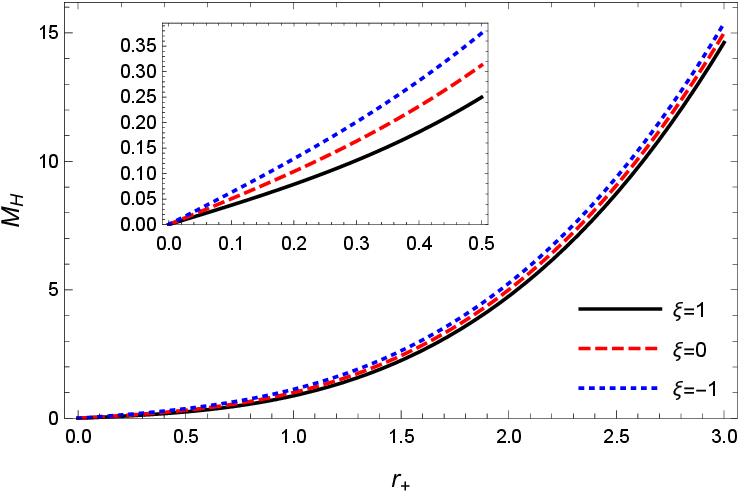} \quad\quad
\includegraphics[width=0.35\textwidth]{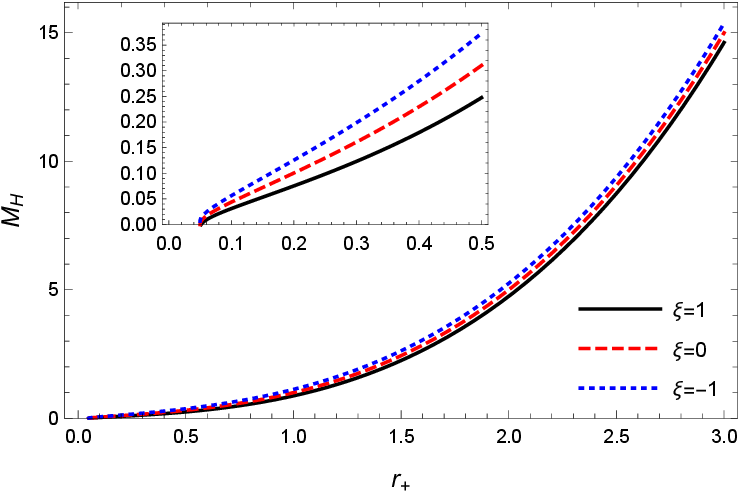}
\caption{ \label{Mplots1} The qualitative representation of the mass function versus the horizon for $P=\frac{3}{8\protect\pi }$ and $\protect\eta =0.1$. In the Left figure $\alpha=0$, and the right figure $\alpha=0.05$.}
\end{figure}

However, using Eq. (\ref{d2}), we can express the BH mass in terms of the event horizon radius $r=r_{+}$ as follows:
\begin{eqnarray}
M_H=\frac{r_{+}}{2} \Bigg\{\sqrt{1-\left(\frac{\alpha}{r_{+}}\right)^2}\Bigg[1+\frac{8\,\pi\, P\, r_{+}^2}{3}\bigg(1-\left(\frac{\alpha}{r_{+}}\right)^2\bigg)\Bigg]-8\,\pi\, \eta^2\, \xi \Bigg\} . \label{mass}   
\end{eqnarray}
Here, it is worth noting that for $\alpha=0=\eta=P$, Eq. (\ref{mass}) reduces to the ordinary Schwarzschild BH mass, while for $\alpha=0=\eta$, it shrinks to the AdS Schwarzschild BH mass \cite{SWH}.  Moreover, only in the absence of quantum correction, Eq. (\ref{mass}) shortens to the AdS Schwarzschild BH mass with global monopoles 
\begin{eqnarray}
M_H=\frac{r_{+}}{2}\Bigg(1-8\,\pi\, \eta^2\, \xi +\frac{8\,\pi\, P\, r_{+}^2}{3}\Bigg) . \label{mass2}   
\end{eqnarray}
Furthermore, for $\eta=0$, it lessens to the quantum corrected-AdS Schwarzschild BH mass \cite{Biz2023}. In Figure \ref{Mplots1} we present the influence of the monopole term on the mass. In particular, we compare its effect in the absence and presence of the quantum correction. 

\begin{center}
\begin{figure}[htb!]
\begin{centering}
\includegraphics[width=0.35\textwidth]{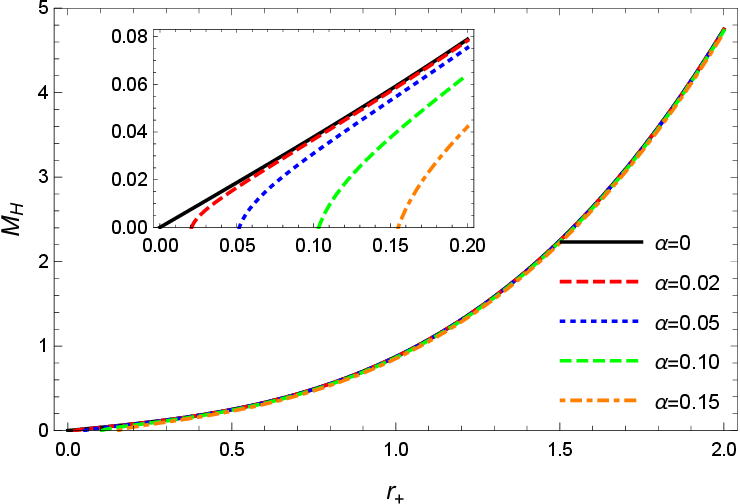}\quad\quad
\includegraphics[width=0.35\textwidth]{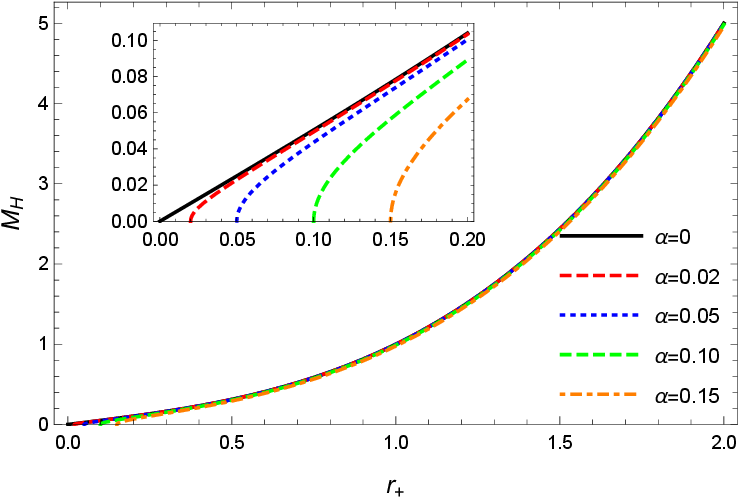}
\par\end{centering}
\hfill\\
\begin{centering}
    \includegraphics[width=0.35\textwidth]{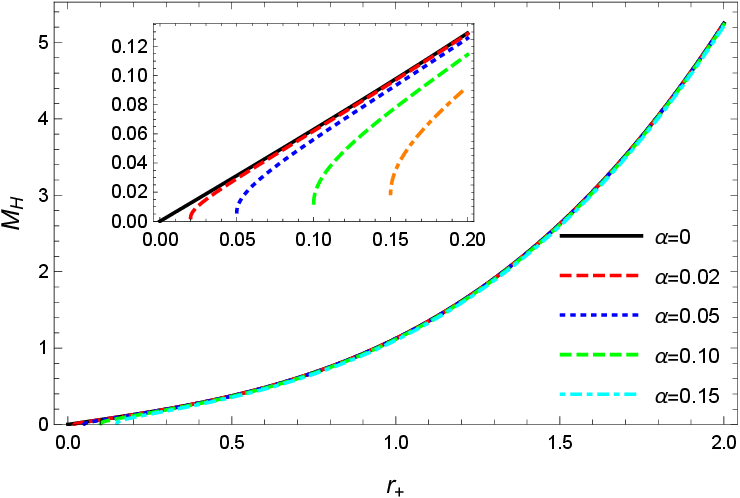}
\par\end{centering}
\caption{\label{Mplots2} The qualitative representation of the mass function versus the horizon in three different monopole scenarios for $P=\frac{3}{8\pi }$ and $\eta =0.1$. In the left figure $\xi=1$, middle one $\xi=0$, and the right one $\xi=-1$.}
\end{figure}
\end{center}

In both scenarios, the BH mass increases for the phantom monopole which was already stated for the non-AdS BH \cite{SCJJ, AHEP}. We also note that quantum corrections do not alter this effect except by setting a lower horizon bound on the mass due to the square root term. Then, we discuss that effect via different values of the correction term for three scenarios in Figure  \ref{Mplots2}.  

All cases show that the quantum effects solely modify the mass on the very small event horizons. { Next, we derive the Hawking temperature  by employing the formula \cite{GM2,GM10,GM11}}
\begin{equation}
T_{H}=\frac{1}{4\pi } \frac{d\mathcal{F}}{dr}\bigg\vert_{r=r_+}.
\end{equation}%
Following the simple derivation, we obtain the Hawking temperature in the form of
\begin{equation}
T_{H}=\frac{1+8 \pi P(r_{+}^{2}-\alpha^2)}{4\pi\sqrt{r_{+}^{2}-\alpha ^{2}}}-\frac{2 \eta ^{2}\xi }{ r_{+}}.\label{tem}
\end{equation}%
Like the mass, the Hawking temperature must have a real, thus, physical value. This fact implies that the event horizon of the BH has to meet the following condition
\begin{equation}
r_{+}>\alpha .
\end{equation}
Before proceeding to the graphical demonstration, we examine Eq. (\ref{tem}) in the $\alpha=\eta=P=0$ and $\alpha=\eta=0$ limits. We find that in both cases the ordinary Schwarzschild Hawking, $T_H=\frac{1}{4\pi r_{+}}$, and the AdS Schwarzschild Hawking temperatures, $T_H=\frac{1}{4\pi r_{+}}+2Pr_{+}$,  recover. Furthermore, in the absence of the quantum correction, Eq. (\ref{tem}) reduces to the AdS Schwarzschild BH mass with monopole
\begin{equation}
    T_H=\frac{1}{4\pi r_{+}}+2Pr_{+}-\frac{2\eta^2 \xi}{r_{+}}, 
\end{equation}
while in the absence of the monopole term, Hawking temperature turns to 
  \begin{equation}
    T_H=\frac{1+8 \pi P(r_{+}^{2}-\alpha^2)}{4\pi\sqrt{r_{+}^{2}-\alpha ^{2}}}.
\end{equation}

\begin{figure}[htb!]
\centering
\includegraphics[width=0.35\textwidth]{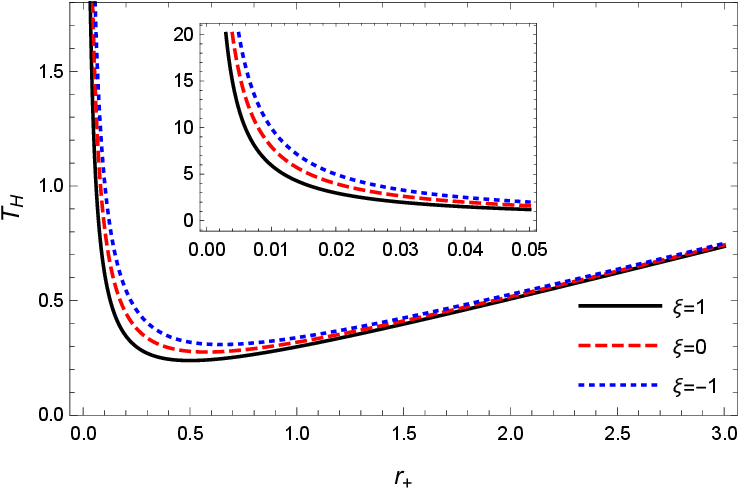}\quad\quad
\includegraphics[width=0.35\textwidth]{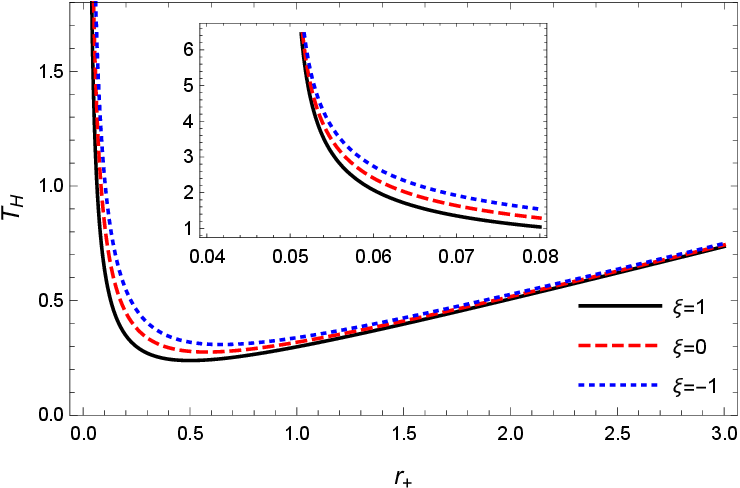}
\caption{ \label{Tplots1} The qualitative representation of the Hawking temperature versus the horizon in three different monopole scenarios for $P=\frac{3}{8\protect\pi }$ and $\protect\eta =0.1$. In the left figure $\alpha=0$  and the right one $\alpha=0.05$.}
\end{figure}

We see that the contribution of the monopole term to the Hawking temperature is articulated as a shift term that decreases or increases in proportion to the inverse of the event horizon. Now, we depict the Hawking temperature versus the event horizon to demonstrate the effects of the monopole and quantum corrections. Figure \ref{Tplots1} shows the case with and without quantum correction corrections.

\begin{center}
\begin{figure}[htb!]
\begin{centering}
\includegraphics[width=0.35\textwidth]{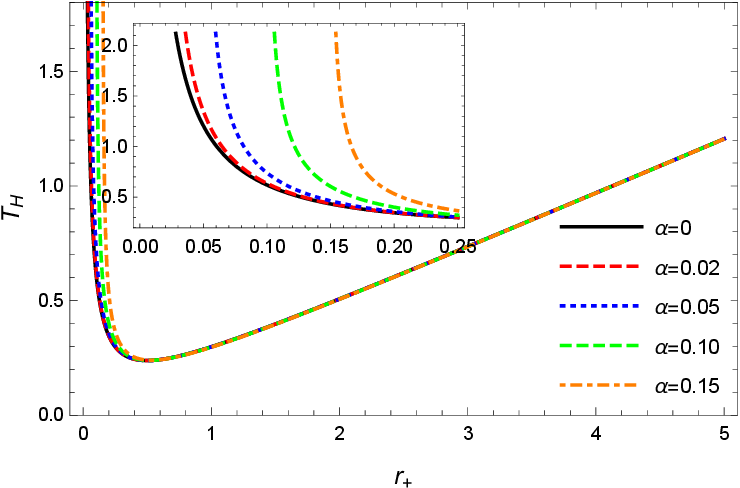}\quad\quad
\includegraphics[width=0.35\textwidth]{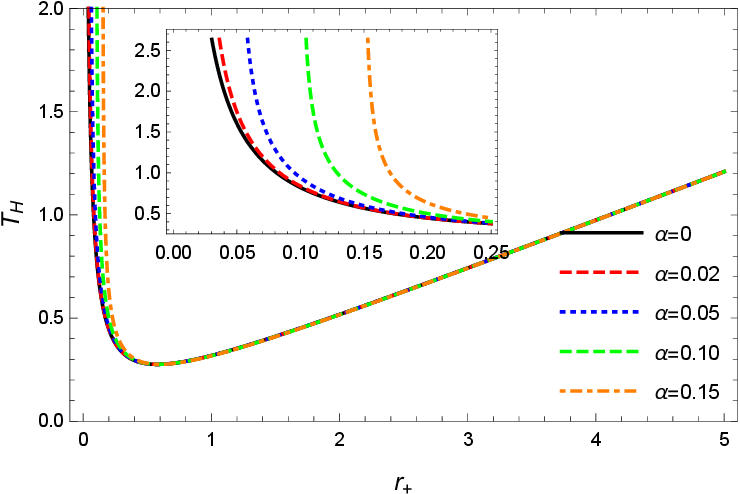}
\par\end{centering}
\hfill\\
\begin{centering}
\includegraphics[width=0.35\textwidth]{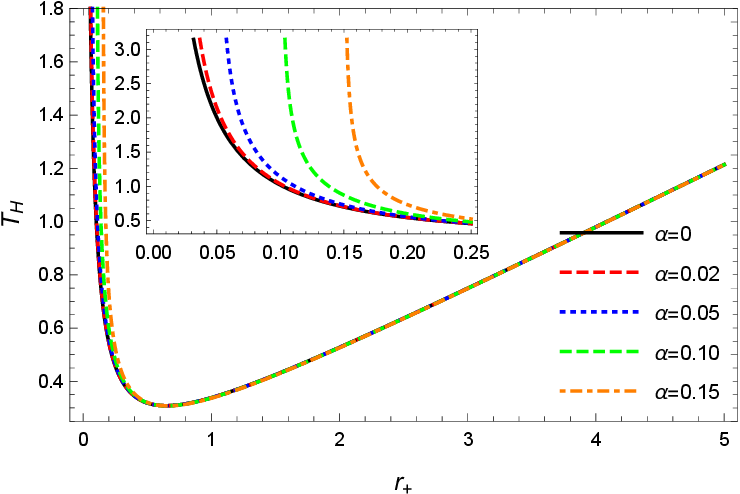}
\par\end{centering}
\caption{\label{Tplots2} The qualitative representation of the Hawking temperature versus
horizon for in three different monopole scenarios $P=\frac{3}{8\protect\pi}$, and $\protect\eta =0.1$. In the left figure  $\xi=1$ , the middle one $\xi=0$, and the right one  $\xi=-1$.}
\end{figure}
\end{center}

In the quantum corrected case, we observe the lower bound of the event horizon. As a common behavior, we see that the Hawking temperature changes sharply in the presence of a monopole $\xi=1$ in both cases. We then compare the impact of the quantum correction magnitude on three scenarios in Figure \ref{Tplots2}.

In all three cases, we see that quantum corrections do not cause any change in large values of the event horizon. We now focus on the heat capacity function. To this end, we employ the following definition \cite{GM2,GM10,GM11}
\begin{equation}
C=\frac{dM}{dT}, \label{heatt}
\end{equation}%
and using Eq. (\ref{heatt}) we obtain the capacity as follows:
\begin{eqnarray}
    C&=& -2\pi r_{+}^{2}  \left(1 -\frac{\alpha^{2}}{r_{+}^{2}}\right)\frac{1+8\pi P ( r_{+}^{2}-\alpha ^{2})-8\pi \eta^2\xi \sqrt{1 -\frac{\alpha^{2}}{r_{+}^{2}}}}{1-8\pi P ( r_{+}^{2}-\alpha ^{2})-8\pi\eta^2\xi \sqrt[3]{ 1 -\frac{\alpha^{2}}{r_{+}^{2}}} }. \label{spheat}
\end{eqnarray}
We notice that for $\alpha=\eta=P=0$,
Eq.(\ref{spheat}) diminishes to the conventional form of the specific heat function of the Schwarzschild BH, $C= -2 \pi r_{+}^2$. Similarly, for $\alpha=\eta=0$, Eq.(\ref{spheat}) gives the heat capacity of the AdS Schwarzschild BH $C= - 2\pi r_{+}^{2} \frac{ 1+8\pi P r_{+}^{2}}{ 1-8\pi P  r_{+}^{2}}.$ 
Moreover, for $\eta=0$, it lessens to the quantum corrected-AdS Schwarzschild BH heat capacity \cite{Biz2023}
    \begin{equation}
        C= -2\pi r_{+}^{2} \left(1 -\frac{\alpha^{2}}{r_{+}^{2}}\right)\frac{  1+8\pi P ( r_{+}^{2}-\alpha ^{2}) }{1-8\pi P ( r_{+}^{2}-\alpha ^{2}) }, 
        \end{equation}
and similarly in the absence of quantum correction  Eq.(\ref{spheat}) reduces to 
    \begin{equation}
C= -2\pi  r_{+}^{2} \frac{ 1+8\pi P  r_{+}^{2}-8\pi \eta^2\xi }{1-8\pi P  r_{+}^{2}-8\pi\eta^2\xi }.
    \end{equation}
Here, we have to point out that the ordinary Schwarzschild BH differs from other scenarios in that it is unstable for all event horizon values. However, in other cases, for example in the AdS Schwarzschild BH scenario, the heat capacity is negative until $r_+=\sqrt{\frac{1}{8\pi P}}$ and then positive. The existing discontinuity in that event horizon value indicates a second-order phase transition. 

\begin{figure}[tbp]
\centering
\includegraphics[width=0.35\textwidth]{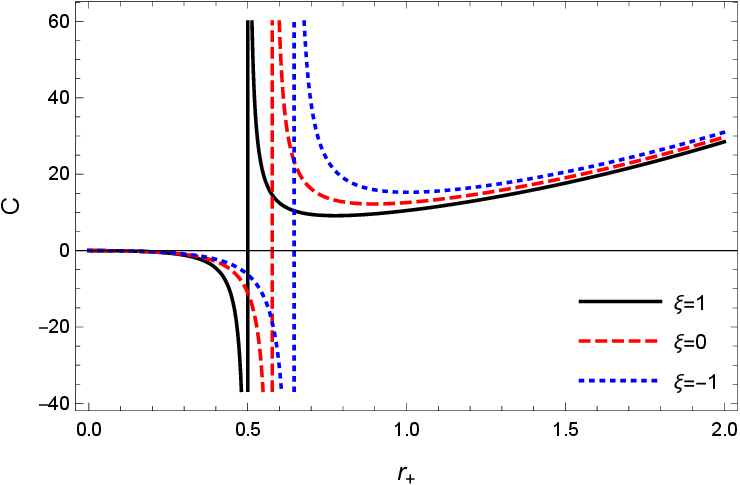}\quad\quad
\includegraphics[width=0.35\textwidth]{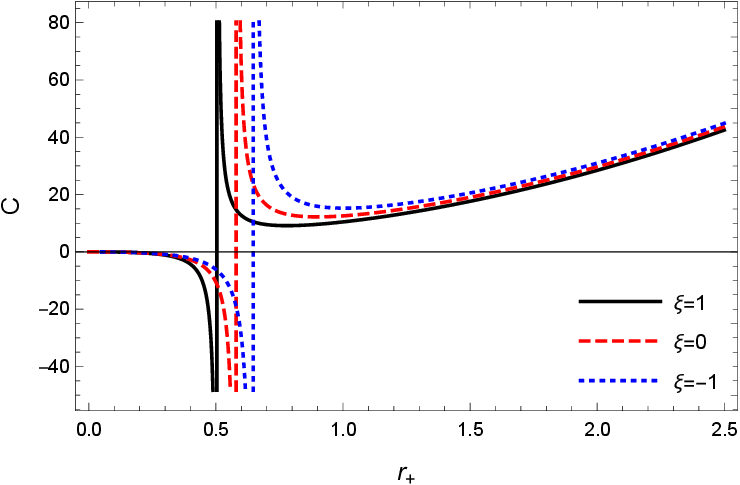}
\caption{\label{Cplots1} The qualitative representation of the heat capacity versus the horizon for $P=\frac{3}{8\protect\pi }$ and $\protect\eta =0.1$. In the left figure $\alpha=0$, and the right one $\alpha=0.5$.}
\end{figure}

Now, we graphically demonstrate the impact of the monopole term on the heat capacity function in Figure \ref{Cplots1}. 

The comparison of Figure \ref{Cplots1} reveals a difference only in very small horizons where the BH is unstable.  On the other hand, we note that both figures indicate the second-order phase transition occurrence in the greater horizon values for the phantom monopole scenarios. 

\begin{figure}[htb!]
\begin{centering}
\includegraphics[width=0.35\textwidth]{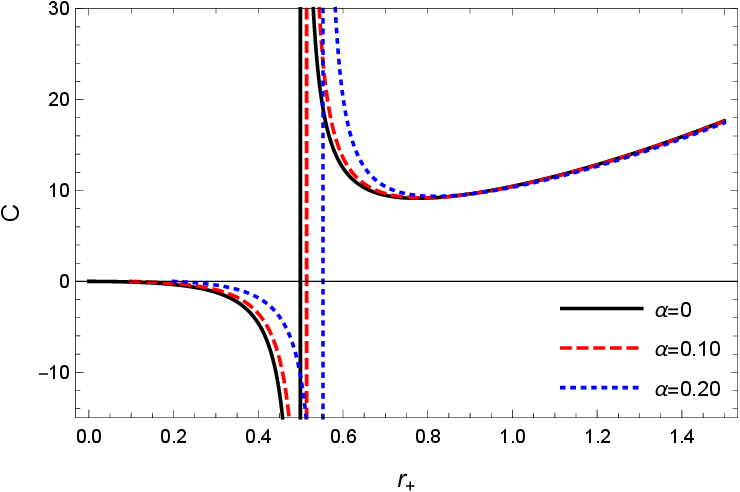}\quad\quad
\includegraphics[width=0.35\textwidth]{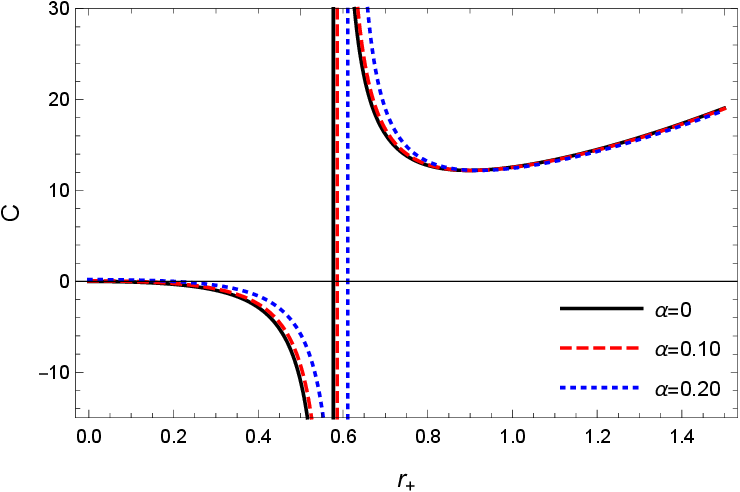}
\par\end{centering}
\begin{centering}
\includegraphics[width=0.35\textwidth]{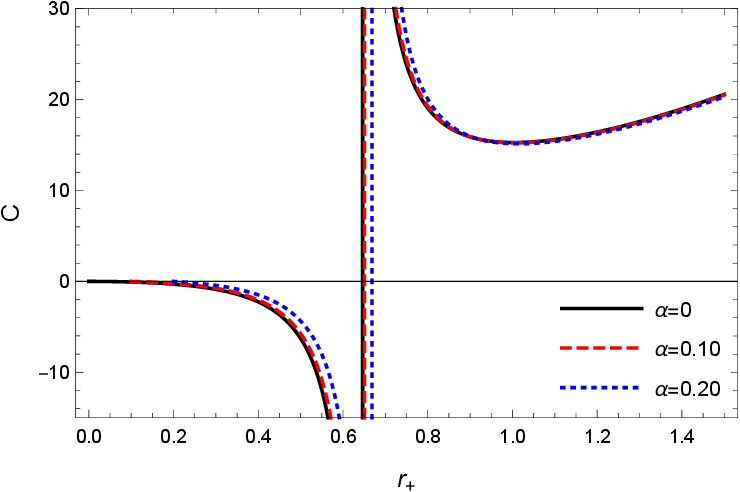}
\par\end{centering}
\caption{\label{Cplots2} The qualitative representation of the heat capacity versus
horizon in three different monopole scenarios for $P=\frac{3}{8\protect\pi}$,  and $\protect\eta =0.1$. In the left figure $\xi=1$ , the middle figure $\xi=0$, and the right one $\xi=-1$.}
\end{figure}

We then display the impact of the quantum correction parameter in Figure \ref{Cplots2}.

We observe that in the greater quantum correction scenarios, the heat capacity function shifts slightly. Next, we derive the BH entropy by employing the customary definition, which is given in the form of
\begin{equation}
    dS= \frac{dM}{T}.\label{ent}
\end{equation}
After the integration, we find the entropy as 
\begin{eqnarray}
   S\simeq \pi r_+^2 =\frac{A}{4}.
\end{eqnarray}
As stated earlier, the event horizon radius $r_{+}$ is related with parameters $\alpha$, and $\xi$ in addition to those $P, M$ given in Eq. (\ref{d2}), thus, entropy and the surface area of the BH are also related with quantum correction term and phantom global monopoles.

\begin{figure}[htb!]
\centering
\includegraphics[width=0.35\textwidth]{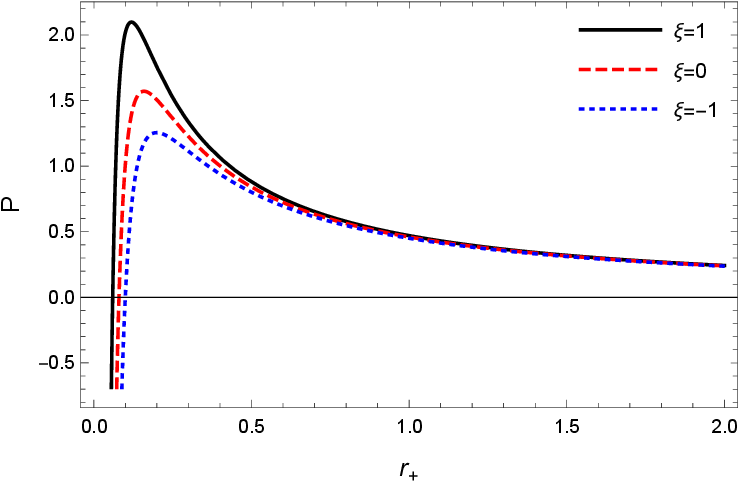}\quad\quad
\includegraphics[width=0.35\textwidth]{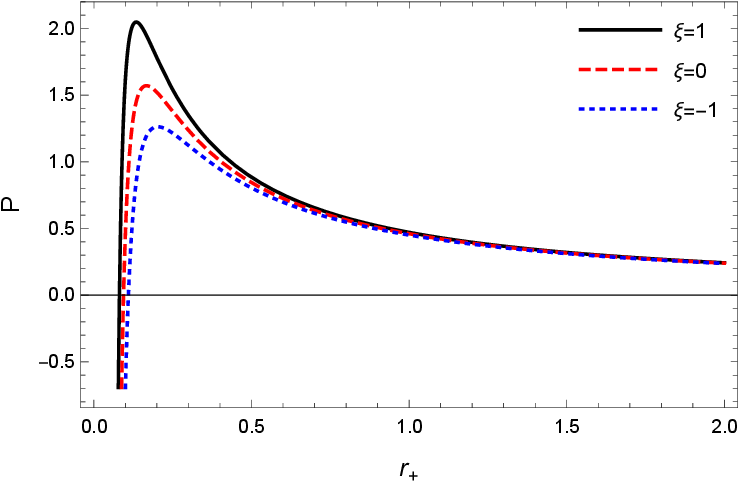}
\caption{\label{Pplots1} The qualitative representation of the pressure isotherms versus the horizon for $T=1$ and $\protect\eta =0.1$. In the left figure $\alpha=0$ and the right one $\alpha=0.5$.}
\end{figure}

Now, we derive the equation of state using  Eq.(\ref{tem}). We find
\begin{equation}
P=-\frac{1}{8\pi
\left( r_{+}^{2}-\alpha ^{2}\right) }+\frac{T_H}{2\sqrt{r_{+}^{2}-\alpha ^{2}}}+\frac{\eta ^{2}\xi }{r_{+}\sqrt{r_{+}^{2}-\alpha ^{2}}}.
\end{equation}%
We then depict pressure isotherms in Figure \ref{Pplots1}, and we compare the effects of the quantum corrections and monopole scenarios. 

\begin{figure}[htb!]
\begin{centering}
\includegraphics[width=0.35\textwidth]{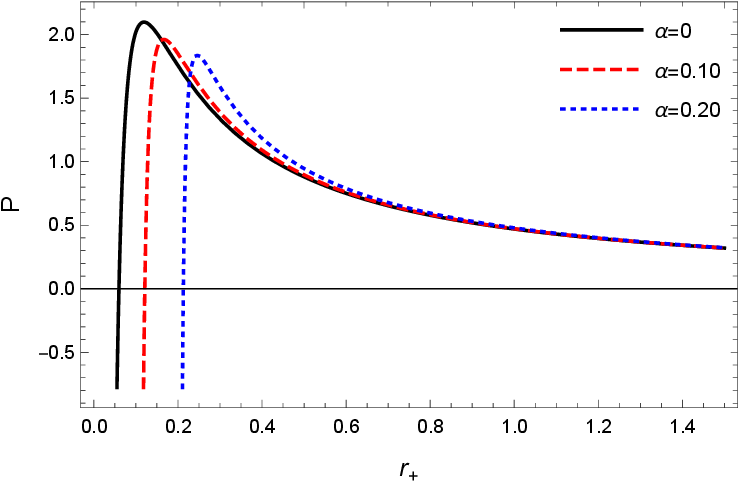}\quad\quad
\includegraphics[width=0.35\textwidth]{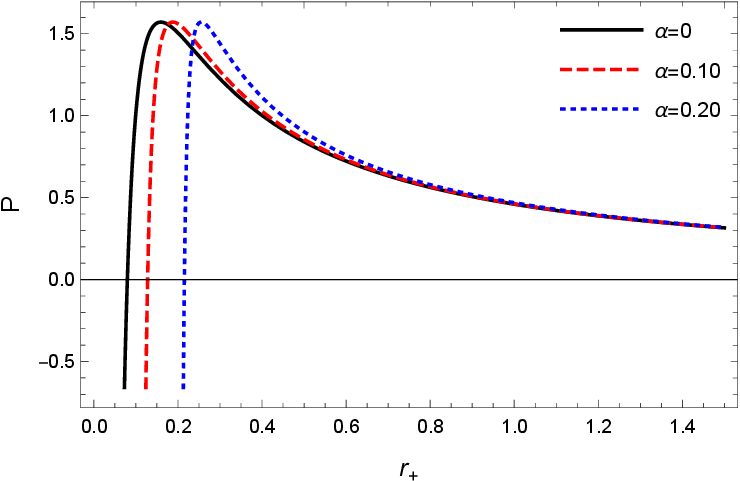}
\par\end{centering}
\hfill\\
\begin{centering}
\includegraphics[width=0.35\textwidth]{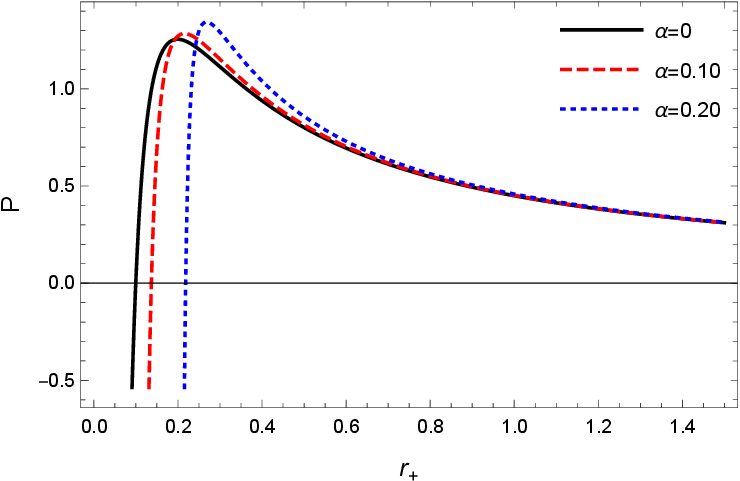}
\par\end{centering}
\caption{\label{Pplots2} The qualitative representation of the pressure isotherms versus horizon in three different monopole scenarios for $T=1$,  and $\eta =0.1$. In the left figure $\xi=1$ , the middle figure $\xi=0$, and the right one $\xi=-1$.}
\end{figure}

Figure \ref{Pplots1} reveal that the pressure isotherms mimic itself including a shift on the event horizon lower bound. Here, we observe that the pressure peak occurs at a smaller pressure value and at a greater horizon in the phantom monopole case. We then depict Figure \ref{Pplots2}.  

We notice that the type of global monopole plays an important role. For example, in the ordinary global monopole case, we observe that as the $\alpha$ value increases, the maximal value of the pressure decreases. However, for the phantom global monopole case, we find that as the $\alpha$ value increases, the maximal value of the pressure also increases. Moreover, we note that if there is no monopole, the maximal value is not affected by the quantum corrections.

\subsection{Effective potential of the system and Shadow behaviors}\label{subsec:2}

In this section, we analyze the effective potential of the system for both massive and massless particles within the chosen BH background (\ref{b}). We investigate this using the Lagrangian method. According to this approach, the effective potential is defined as \cite{FA, FA2, FA3, FA4, FA5, FA6,FA7, BH2024,GM8,GM9,GM10} 
\begin{equation}
\mathcal{L}=\frac{1}{2}\,g_{\mu\nu}\,\dot{x}^{\mu}\,\dot{x}^{\nu},
\label{b2}
\end{equation}
where dot represents ordinary derivative w. r. t an affine parameter of the
curve, $\tau$, and $g_{\mu\nu}$ is the metric tensor for the line-element (\ref{b}).

Using the line-element (\ref{b}) in the equatorial plane $\theta=\frac{\pi}{2%
}$, we obtain the following expression
\begin{equation}
\mathcal{L}=\frac{1}{2}\,\Bigg[-\mathcal{F}(r)\,\dot{t}^2+\mathcal{F}^{-1}
(r)\,\dot{r}^2+r^2\,\dot{\phi}^2\Bigg].  \label{b3}
\end{equation}
One can see that the metric tensor $g_{\mu\nu}$ for the considered
space-time is independent of $(t, \phi)$. Therefore, there are two constants
of motions associated with these coordinates given by 
\begin{eqnarray}
&&-E=-\mathcal{F}(r)\,\dot{t} \Rightarrow \dot{t}=\frac{E}{\mathcal{F}(r)},
\label{b4} \\
&&L=r^2\,\dot{\phi}\Rightarrow \dot{\phi}=\frac{L}{r^2},  \label{b5}
\end{eqnarray}
where $E$ is the conserved energy parameter, and $L$ is the conserved $z$%
-component of the angular momentum.

With these, the Lagrangian (\ref{b3}) for light-like or time-like geodesics
becomes 
\begin{equation}
\Big(\frac{dr}{d\tau}\Big)^2+\mathcal{F}(r)\,\Big(-\epsilon+\frac{L^2}{r^2}%
\Big)=E^2,  \label{b6}
\end{equation}
where $\epsilon=0$ for null geodesics and $-1$ for time-like geodesics. Eq. (\ref{b6}) can be seen as describing the dynamics of a classical
particle of energy E subject to an effective potential given by 
\begin{equation}
V_{eff} (r)=\Bigg[\frac{1}{r}\,\sqrt{r^2-\alpha^2}-\frac{2\,M}{r}%
-8\,\pi\,\eta^2\,\xi+\frac{(r^2-\alpha^2)^{3/2}}{r\,\ell^2}\Bigg]\,\Big(%
-\epsilon+\frac{L^2}{r^2}\Big).  \label{b7}
\end{equation}

\begin{center}
\begin{figure}[htb!]
\begin{centering}
\includegraphics[width=0.35\textwidth]{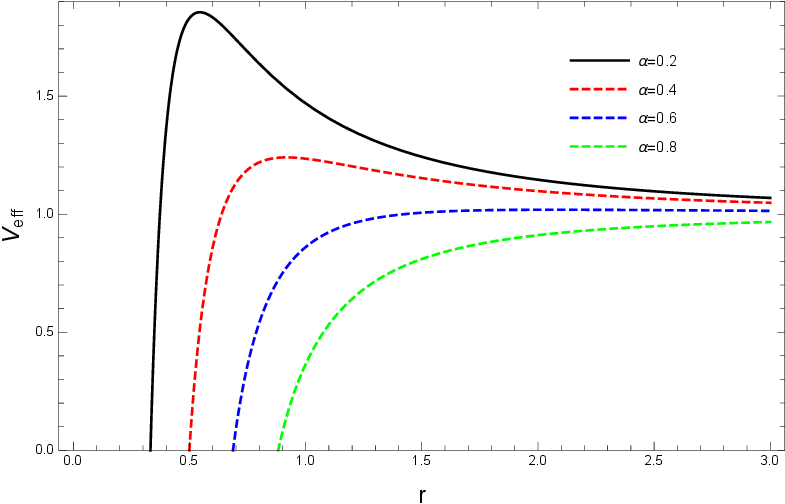}\quad\quad
\includegraphics[width=0.35\textwidth]{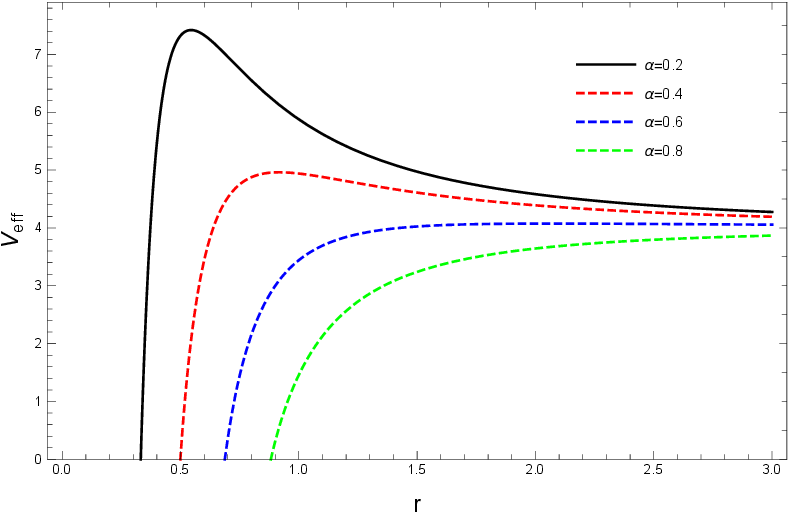}
\par\end{centering}
\hfill\\
\begin{centering}
\includegraphics[width=0.35\textwidth]{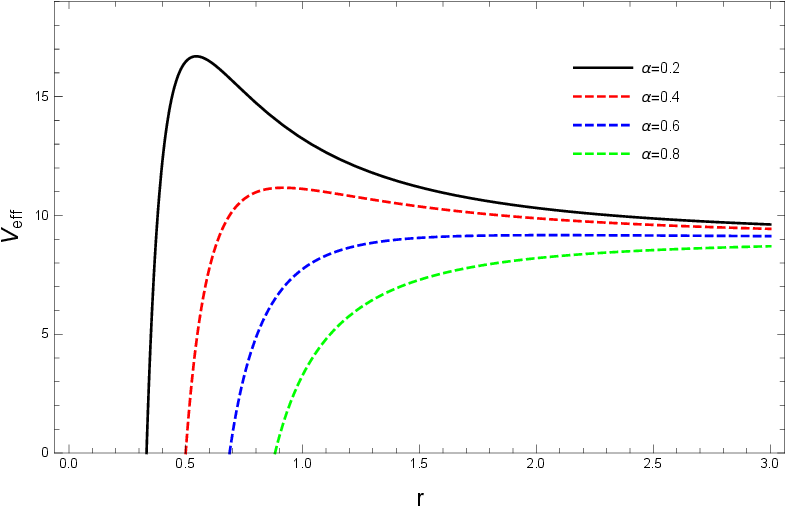}
\par\end{centering}
\caption{\label{Vppotentel} The effective potential for null geodesics vs. $r$ for different
values of $\protect\alpha$. Here $\protect\xi=1$. In the left figure $L=1$ , the middle figure $L=2$, and the right one $L=3$.} 
\end{figure}
\end{center}

\begin{center}
\begin{figure}[htb!]
\begin{centering}
\includegraphics[width=0.35\textwidth]{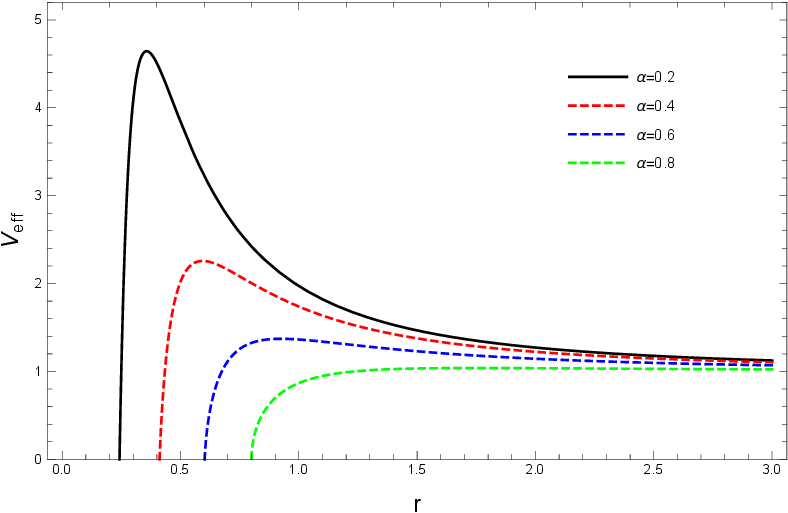}\quad\quad
\includegraphics[width=0.35\textwidth]{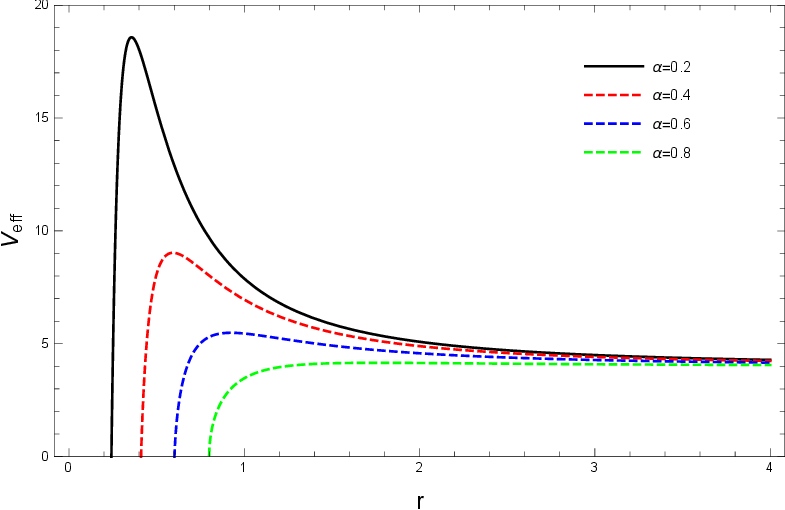}
\par\end{centering}
\hfill\\
\begin{centering}
\includegraphics[width=0.35\textwidth]{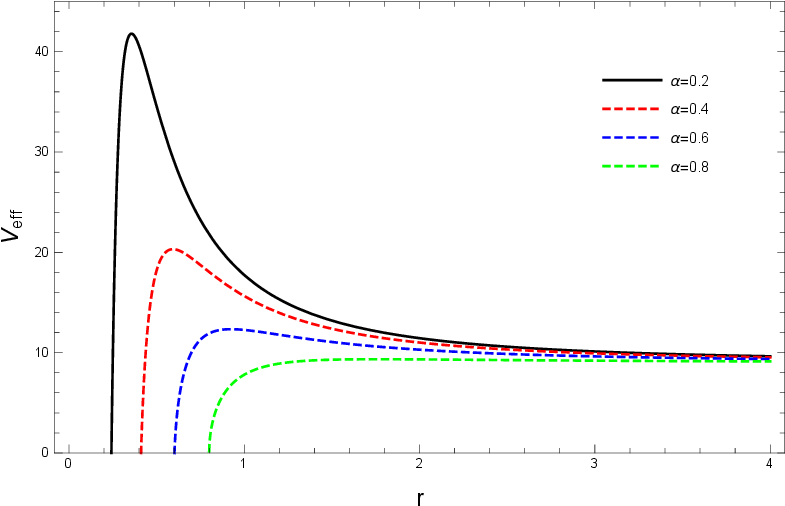}
\par\end{centering}
\caption{\label{Vnpotentel} The effective potential for null geodesics vs. $r$ for different
value of $\protect\alpha$. Here $\xi=-1$. In the left figure $L=1$ , the middle figure $L=2$, and the right one $L=3$.}
\end{figure}
\end{center}

\begin{center}
\begin{figure}[tbh]
\begin{centering}
\includegraphics[width=0.35\textwidth]{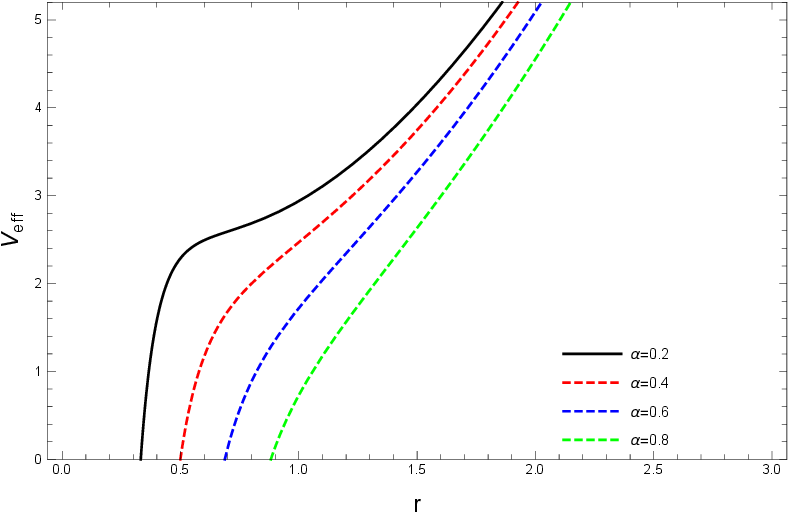}\quad\quad
\includegraphics[width=0.35\textwidth]{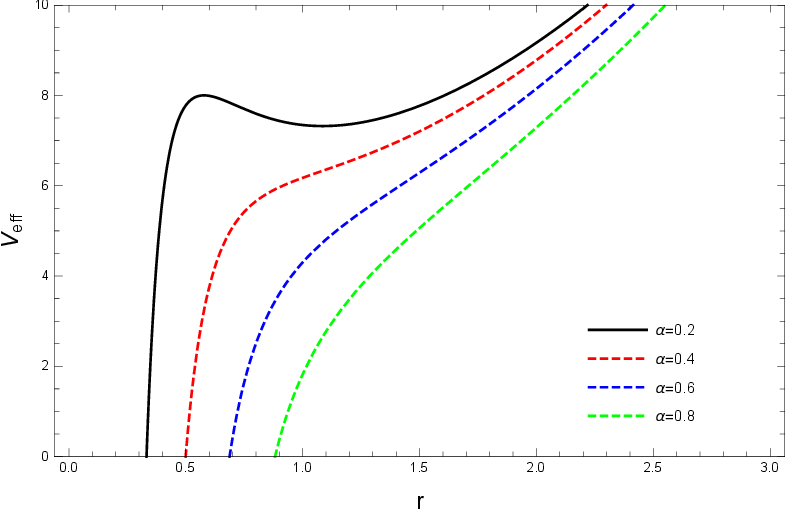}
\par\end{centering}
\hfill\\
\begin{centering}
\includegraphics[width=0.35\textwidth]{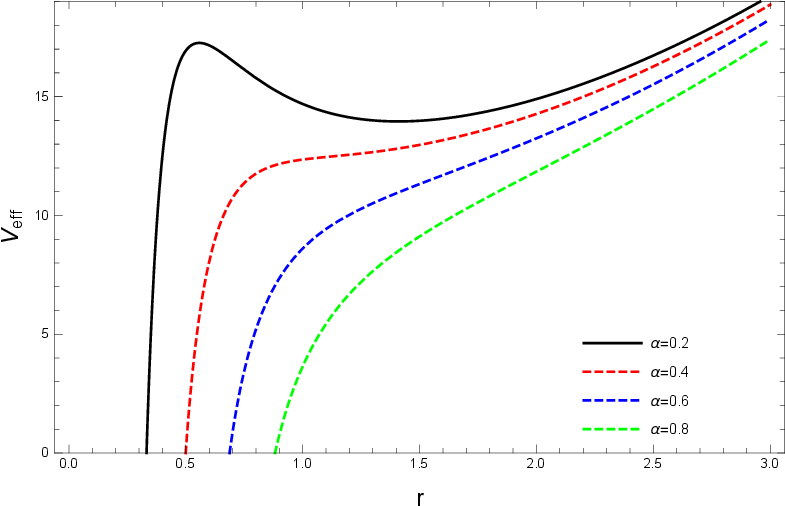}
\par\end{centering}
\caption{\label{Vpotett} The effective potential for time-like geodesics versus event horizon for
different values of $\protect\alpha $ for the usual global monopole case. In the left figure $L=1$ , the middle figure $L=2$, and the right one $L=3$.}
\end{figure}
\end{center}

\begin{center}
\begin{figure}[htb!]
\begin{centering}
\includegraphics[width=0.35\textwidth]{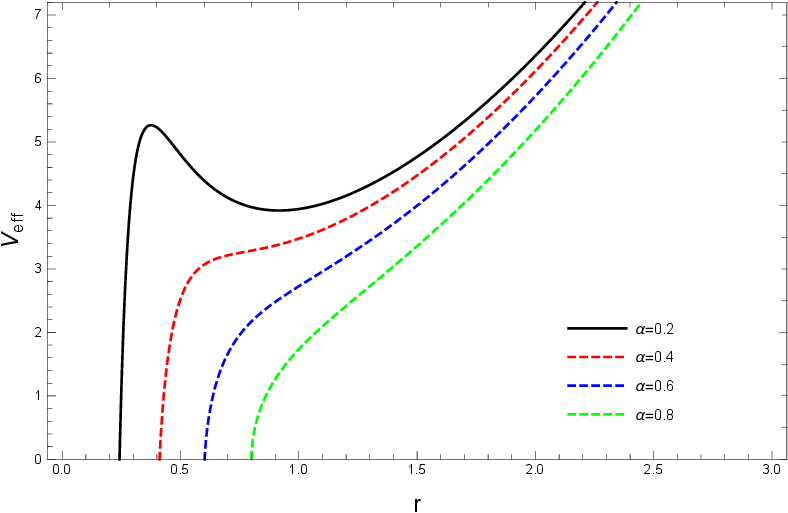}\quad\quad
\includegraphics[width=0.35\textwidth]{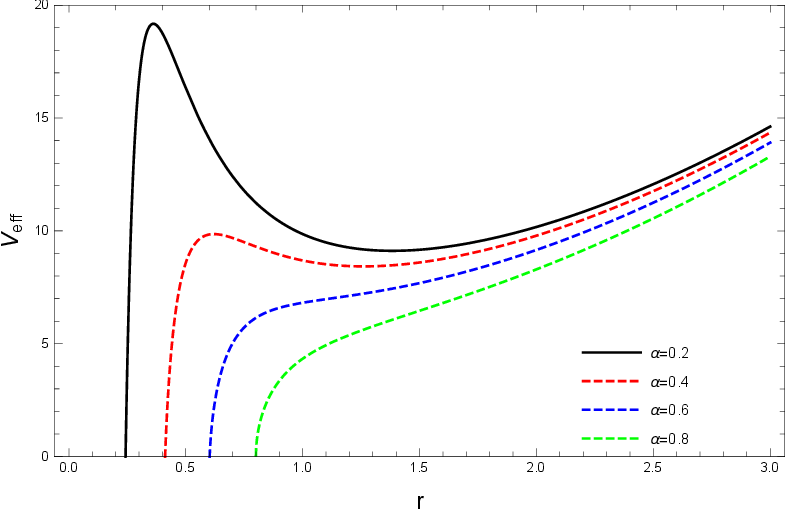}
\par\end{centering}
\hfill\\
\begin{centering}
\includegraphics[width=0.35\textwidth]{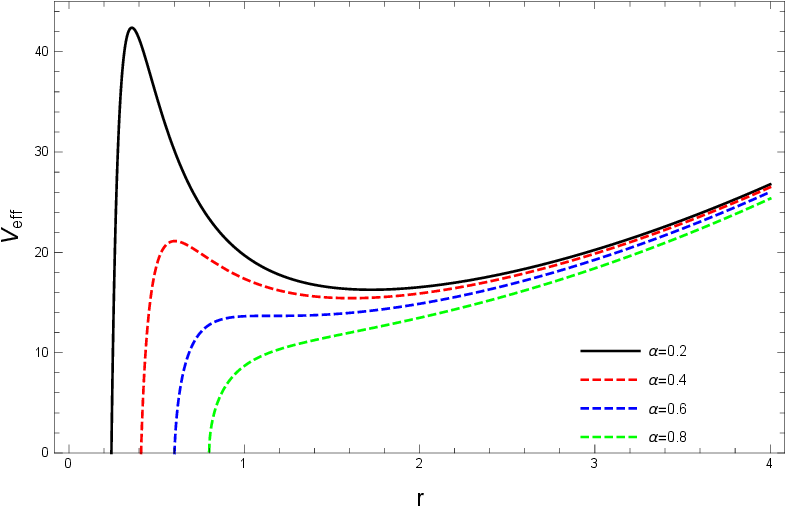}
\par\end{centering}
\caption{\label{Vnnpotentel} The effective potential for time-like geodesics versus event horizon for different values of $\protect\alpha $ for the phantom global monopole case. In the left figure $L=1$ , the middle figure $L=2$, and the right one $L=3$.}
\end{figure}
\end{center}

We have produced several graphs depicting the effective potential for null and time-like geodesics, factoring in both ordinary and phantom global monopoles across various values of the correction parameter $\alpha$. In Figure \ref{Vppotentel}, we plotted the effective potential for null geodesics considering the ordinary global monopole, while Figure \ref{Vnpotentel} showcases the same for phantom global monopoles. We varied the angular momentum $L$ across values of $1$, $2$, and $3$ to observe its impact. Similarly, Figures \ref{Vpotett} and \ref{Vnnpotentel} exhibit the effective potential for time-like geodesics. Again, we've explored different values of $\alpha$ while considering both ordinary and phantom global
monopoles. In Figures \ref{Vppotentel} and \ref{Vnpotentel}, we have set $M=0.1$, $P=\frac{3}{8\pi }$, $\eta =0.1$, and $\epsilon =0$, while in Figures  \ref{Vpotett} and \ref{Vnnpotentel}, we choose $M=0.1,$ $P=\frac{3}{8\pi }$, $\eta =0.1$ and $\epsilon=-1$.

To examine the equation governing null geodesics in the spacetime of a
deformed AdS BH with phantom global monopoles and analyzed how the Quantum
correction affects the evolution of photons, we employ the Hamilton-Jacobi
action,%
\begin{equation}
\frac{\partial S}{\partial \tau }=-\frac{1}{2}g^{\mu \nu }\frac{\partial S}{%
\partial x^{\mu }}\frac{\partial S}{\partial x^{\nu }},  \label{hj}
\end{equation}%
where $S$ is the Jacobi action.To examine the Hamilton-Jacobi equation, we
adopt an ansatz in the following format:%
\begin{equation}
S=\frac{m^{2}}{2}\tau -Et+L\phi +S_{\theta }\left( \theta \right)
+S_{r}\left( r\right) ,  \label{ss}
\end{equation}%
Here we set $m=\epsilon =0$ for a photon. Using the conserved quantities $E$
and $L$ and inserting Eq. (\ref{ss}) into the Hamilton-Jacobi equation, Eq. (\ref%
{hj}), we find

\begin{equation}
r^{2}\left( \frac{\partial S_{r}}{\partial r}\right) ^{2}=\frac{r^{2}E^{2}}{%
\mathcal{F}^{2}(r)}-\frac{\left( L^{2}+\mathcal{K}\right) }{\mathcal{F}(r)\,}%
.
\end{equation}%
\begin{equation}
\left( \frac{\partial S_{\theta }}{\partial \theta }\right) ^{2}=\mathcal{K}%
-L^{2}\cot ^{2}\theta .
\end{equation}%
where $\mathcal{K}$ is called Carter's constant. By employing the subsequent
definition of canonically conjugate momentum,

\begin{equation}
\frac{\partial S_{\theta }}{\partial \theta }=p_{\theta }=r^{2}\frac{%
\partial \theta }{\partial \tau },  \label{teta}
\end{equation}%
and%
\begin{equation}
\frac{\partial S_{r}}{\partial r}=p_{r}=\frac{1}{\mathcal{F}(r)}\frac{%
\partial r}{\partial \tau },  \label{eqr}
\end{equation}
we can obtain the null geodesic equations as follows:%
\begin{equation}
r^{2}\frac{\partial \theta }{\partial \tau }=\pm \sqrt{\Theta }=\sqrt{%
\mathcal{K}-L^{2}\cot ^{2}\theta },
\end{equation}%
\begin{equation}
r^{2}\dot{r}=\pm \sqrt{\mathcal{R}}=\sqrt{r^{4}E^{2}-r^{2}\mathcal{F}%
(r)\left( L^{2}+\mathcal{K}\right) }.  \label{rad}
\end{equation}%
Now, our attention shifts towards the radial equation as we introduce the
potential.%
\begin{equation}
\left( \frac{dr}{d\tau }\right) ^{2}+V\left( r\right) =0,
\end{equation}%
where%
\begin{equation}
V\left( r\right) =\left( L^{2}+\mathcal{K}\right) \frac{\mathcal{F}(r)}{r^{2}%
}-E^{2}.
\end{equation}
To discover the unstable circular orbits, we impose the following conditions 
\begin{equation}
\left. V\left( r\right) \right\vert _{r=r_{p}}=0,  \label{cd}
\end{equation}%
\begin{equation}
\left. \frac{\partial V\left( r\right) }{\partial r}\right\vert _{r=r_{p}}=0,
\label{scd}
\end{equation}%
and check that $\left. \frac{\partial ^{2}V\left( r\right) }{\partial r^{2}}%
\right\vert _{r=r_{p}}<0$, where $r_{p}$ is the radius of the photon sphere.
Using the condition (\ref{cd}), leads%
\begin{equation}
\beta +\delta ^{2}=\frac{r_{p}^{2}}{\frac{1}{r_{p}}\,\sqrt{r_{p}^{2}-\alpha
^{2}}-\frac{2\,M}{r_{p}}-8\,\pi \,\eta ^{2}\,\xi +\frac{(r_{p}^{2}-\alpha
^{2})^{3/2}}{r_{p}\,\ell ^{2}}},  \label{beta}
\end{equation}%
where we have applied the definitions of Chandrasekhar constants:%
\begin{equation}
\beta =\frac{\mathcal{K}}{E^{2}}\text{ and }\delta =\frac{L}{E}.
\end{equation}
The second condition (\ref{scd}) leads to%
\begin{equation}
r_{p}\mathcal{F}^{\prime }\left( r_{p}\right) -2\mathcal{F}\left(
r_{p}\right) =0,
\end{equation}
or%
\begin{equation}
\frac{3\alpha ^{2}+6M\sqrt{r_{p}^{2}-\alpha ^{2}}-8\pi \alpha
^{4}P+r_{p}^{2}\left( 8\pi \alpha ^{2}P-2\right) +16\pi \eta ^{2}\xi r_{p}%
\sqrt{r_{p}^{2}-\alpha ^{2}}}{r_{p}\sqrt{r_{p}^{2}-\alpha ^{2}}}=0.
\label{ppsph}
\end{equation}

In this scenario, obtaining an exact solution for Eq. (\ref{ppsph}) is not
feasible. Consequently, we opt for numerical methods for solving it. The
presence of quantum correction and phantom global monopoles introduces
3-additional parameter, $\alpha ,$ $\eta $ and $\xi $, in Eq. (\ref{ppsph}).
We explore various values for $\alpha,$ $\eta $ and $\xi $ and determine the
photon sphere radius, $r_{p}$, by numerically solving Eq. (\ref{ppsph}).
Subsequently, we calculate the values of $\beta +\delta ^{2}$\ in Tables \ref%
{tab1} and \ref{tab2}. 

\begin{table}[tbp]
\center%
\begin{tabular}{l|ll|ll}
\hline
& \multicolumn{2}{|c|}{$\eta =0.1$} & \multicolumn{2}{|c}{$\eta =0.01$} \\ 
\hline
$\alpha $ & $r_{p}$ & $\beta +\delta ^{2}$ & $r_{p}$ & $\beta +\delta ^{2}$
\\ \hline\hline
$10^{-1}$ & 4.09232 & 0.985601 & 3.05678 & 0.96612 \\ 
$5\times 10^{-2}$ & 4.0281 & 0.984926 & 3.01974 & 0.964943 \\ 
$10^{-2}$ & 4.00793 & 0.984705 & 3.00804 & 0.964561 \\ 
$5\times 10^{-3}$ & 4.0073 & 0.984698 & 3.00768 & 0.964549 \\ 
$10^{-3}$ & 4.0071 & 0.984696 & 3.00756 & 0.964545 \\ \hline\hline
\end{tabular}%
\caption{The values of photon radius, $r_{p}$, and impact parameters, for
different values of $\protect\alpha $. $\ $We use $M=1,$ $P=\frac{3}{8%
\protect\pi }$ and $\protect\xi =1$.}
\label{tab1}
\end{table}

\begin{table}[tbp]
\center%
\begin{tabular}{l|ll|ll}
\hline
& \multicolumn{2}{|c|}{$\eta =0.1$} & \multicolumn{2}{|c}{$\eta =0.01$} \\ 
\hline
$\alpha $ & $r_{p}$ & $\beta +\delta ^{2}$ & $r_{p}$ & $\beta +\delta ^{2}$
\\ \hline\hline
$10^{-1}$ & 2.42985 & 0.934713 & 3.04124 & 0.965616 \\ 
$5\times 10^{-2}$ & 2.40549 & 0.932939 & 3.00455 & 0.964428 \\ 
$10^{-2}$ & 2.39777 & 0.932365 & 2.99296 & 0.964042 \\ 
$5\times 10^{-3}$ & 2.39753 & 0.932347 & 2.9926 & 0.964030 \\ 
$10^{-3}$ & 2.39746 & 0.932341 & 2.99248 & 0.964026 \\ \hline\hline
\end{tabular}%
\caption{The values of photon radius, $r_{p}$, and impact parameters, for
different values of $\protect\alpha $. $\ $We use $M=1,$ $P=\frac{3}{8%
\protect\pi }$ and $\protect\xi =-1$.}
\label{tab2}
\end{table}

To find the shape of the BH, we define and employ the following
celestial coordinate system

\begin{equation}
X=\lim_{r_{o}\rightarrow \infty }\left( -r_{o}^{2}\sin \theta _{o}\frac{%
d\phi }{dr}\right) ,
\end{equation}%
\begin{equation}
Y=\lim_{r_{o}\rightarrow \infty }\left( r_{o}^{2}\frac{d\theta }{dr}\right) .
\end{equation}

\begin{figure}[htb!]
\centering
\includegraphics[width=0.35\textwidth]{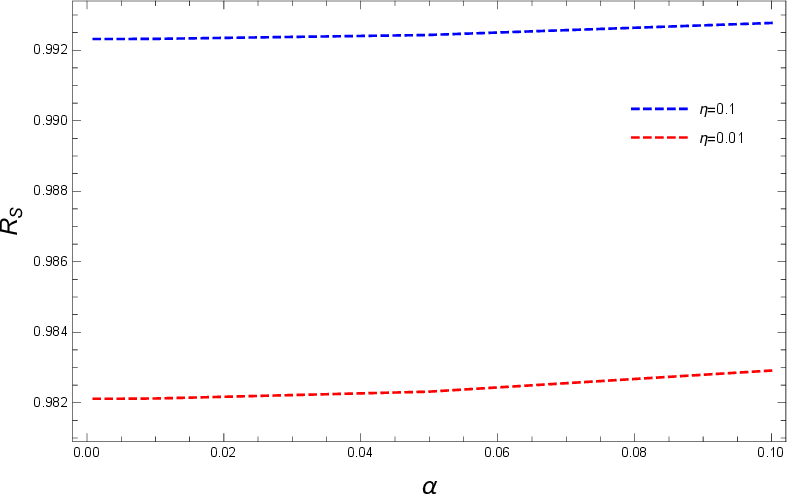}\quad\quad
\includegraphics[width=0.35\textwidth]{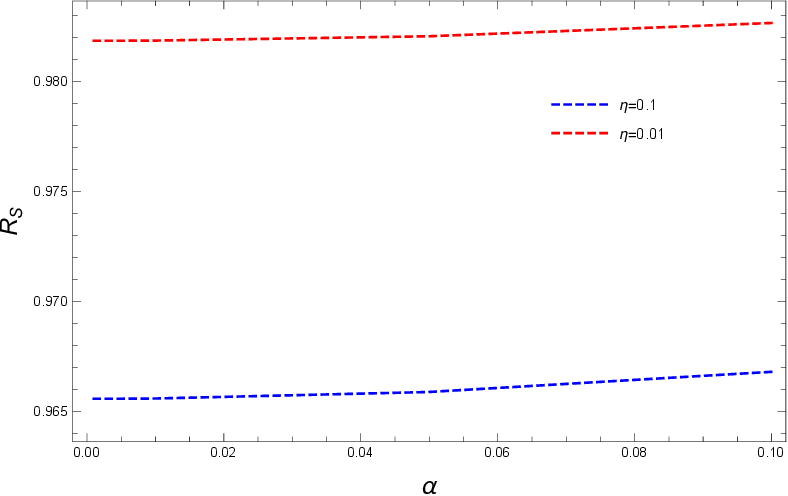}
\caption{\label{raplots1} Variation of the radius of BH shadow. The left figure for the ordinary global monopole and the right one for Phantom global monopole.}
\end{figure}

Here $r_{o}$ represents the distance from the BH to the observer's, $%
\theta _{o}$, called the inclination angle, is the angle between the Black
Hole's rotation axis and the observer's line of sight, and $X$ measures the
apparent perpendicular distance of the shadow from the symmetry axis, as
seen by the observer, $Y$ measures the apparent perpendicular distance of
the shadow itself, as seen by the observer. By utilizing the equations that
describe the null geodesics, we can derive the relationships between the
celestial coordinate system and the impact parameters $\delta $ and $\beta $%
, which characterize the light ray trajectories. These relationships are
expressed as follows:%
\begin{equation}
X=-\frac{\delta }{\sin \theta _{o}},
\end{equation}%
\begin{equation}
Y=\pm \sqrt{\beta -\delta ^{2}\cot ^{2}\theta _{o}}.
\end{equation}%
In the equatorial plane ($\theta _{o}=\pi /2$ ), $X$ and $Y$ becomes%
\begin{equation}
X=\delta \text{ and }Y=\sqrt{\beta }.
\end{equation}%
As a result, the Eq. (\ref{beta}) can be rewritten in the following
form:

\begin{equation}
X^{2}+Y^{2}=\beta +\delta ^{2}=\frac{r_{p}^{2}}{\frac{1}{r_{p}}\,\sqrt{%
r_{p}^{2}-\alpha ^{2}}-\frac{2\,M}{r_{p}}-8\,\pi \,\eta ^{2}\,\xi +\frac{%
(r_{p}^{2}-\alpha ^{2})^{3/2}}{r_{p}\,\ell ^{2}}}=R_{s}^{2},
\end{equation}
where $R_{s}$  represents the radius of the BH's shadow. Figure \ref{raplots1} shows the variation of the shadow with quantum correction $\alpha$, for two values of $\eta$. The plots are shown for both $\xi=+1,-1$ It is observed that $R_{s}$ increases with the increase in quantum correction. 

\bigskip 

\subsection{Energy Emission Rate}\label{subsec:3}

\begin{figure}[htb!]
\centering
\includegraphics[width=0.35\textwidth]{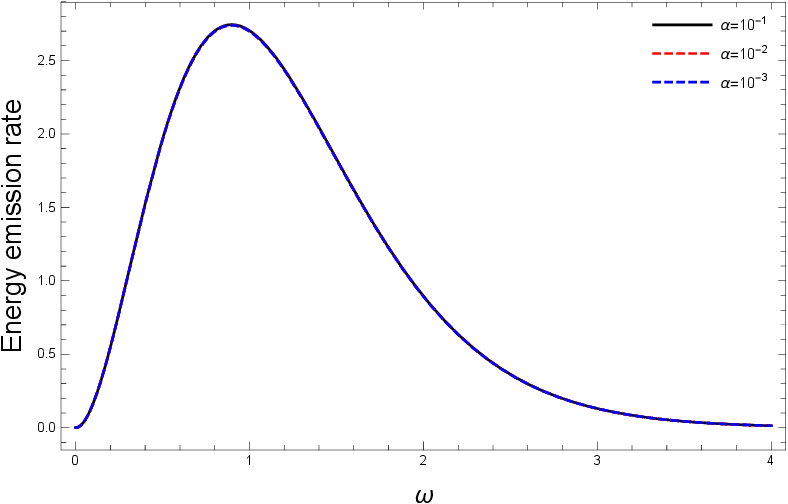}\quad\quad
\includegraphics[width=0.35\textwidth]{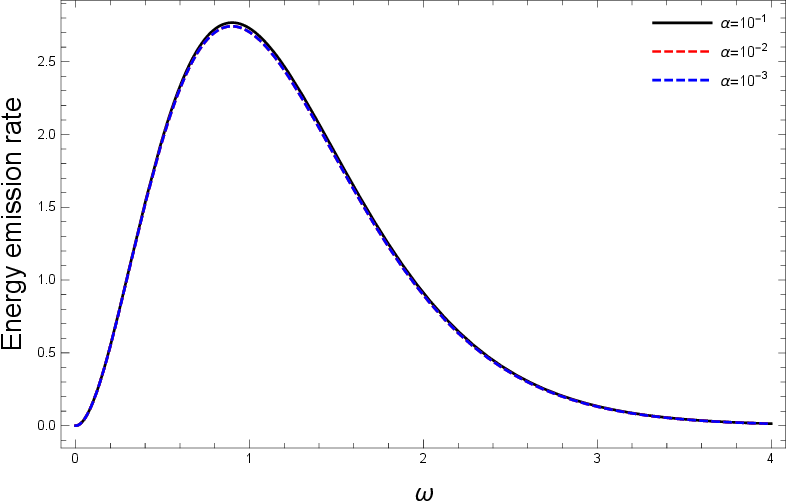}
\caption{\label{EMplots1} Energy emission rate behaviors for various values of the $\alpha$ parameter and $\eta=0.01$. In the left figure $\xi=1$ and he right one $\xi=-1$.}
\end{figure}

\begin{center}
\begin{figure}[htb!]
\begin{centering}
\includegraphics[width=0.35\textwidth]{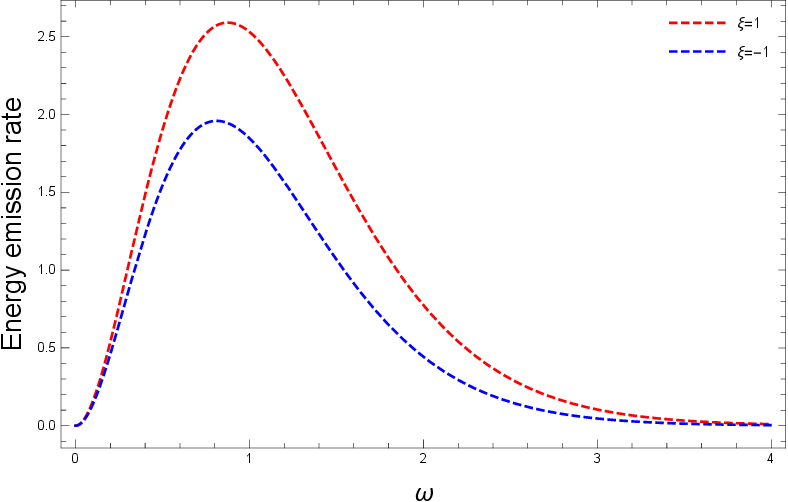}\quad\quad
\includegraphics[width=0.35\textwidth]{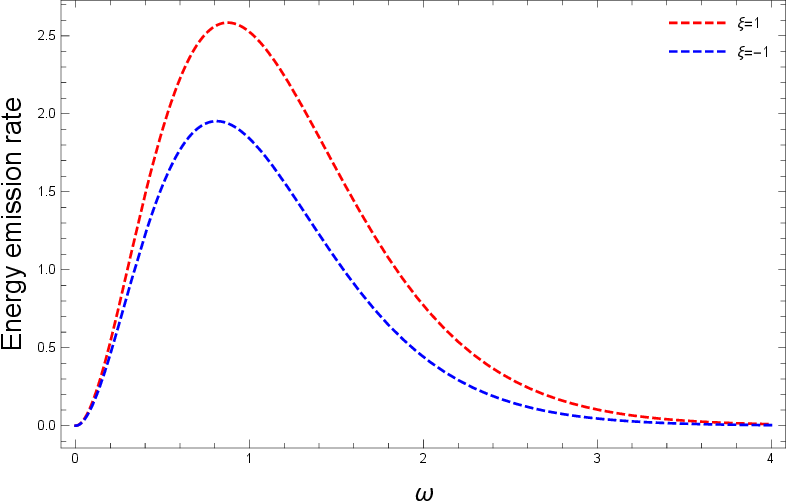}
\par\end{centering}
\hfill\\
\begin{centering}
\includegraphics[width=0.35\textwidth]{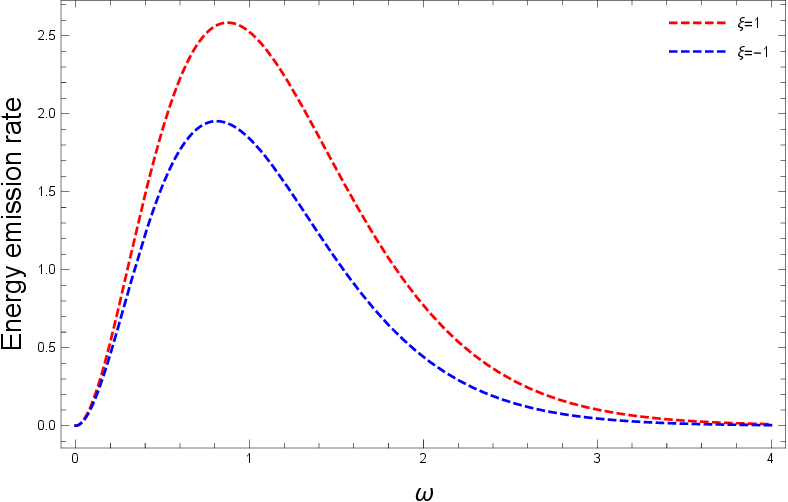}
\par\end{centering}
\caption{\label{EMplots2} Energy emission rate behaviors for various values of the $\xi$ parameter and $\eta=0.1$. In the left figure $\alpha=0.1$ , the middle figure $\alpha=0.01$, and the right one $\alpha=0.001$.}
\end{figure}
\end{center}

It is known that, near the horizon quantum fluctuations lead to the
creation and annihilation of particles. Through this process, particles with
positive energy can tunnel out and escape from the BH's interior
region, resulting in what is known as Hawking radiation. This Hawking
radiation causes BH to evaporate over time. In this section, we
examine the associated rate of energy emission. For an observer located far
away, the high-energy absorption cross-section approaches the BH's
shadow. The absorption cross-section of the BH oscillates and
converges to a limiting constant value $\sigma _{\lim }$ at high energies.
This limiting constant value turns out to be approximately equal to the area
of the photon sphere and can be expressed as%
\begin{equation}
\frac{d^{2}E}{dtd\omega }=\frac{2\pi ^{2}\sigma _{\lim }}{e^{\frac{\omega }{%
T_{H}}}-1}\omega ^{3}.  \label{enem}
\end{equation}%
Here $\omega $ represent the emission frequency,  $T_{H}$ is the Hawking
temperature and $\sigma _{\lim }$ is the absorption cross-section, which is
approximately equal to the geometrical cross-sectional area of the photon
sphere%
\begin{equation}
\sigma _{\lim }\simeq \pi R_{s}^{2}.  \label{set}
\end{equation}
Employing Eq. (\ref{set}) into Eq. (\ref{enem}), we derive the black hole emission energy rate as%
\begin{equation}
\frac{d^{2}E}{dtd\omega }=\frac{2\pi ^{3}R_{s}^{2}}{e^{\frac{\omega }{T_{H}}%
}-1}\omega ^{3}.
\end{equation}
Figure (\ref{EMplots1})  show the variation of energy emission rate with frequency $\omega$ for 3 values of $\alpha$. The plots are shown for both $\xi=+1,-1$. We observe that all curves are coincide for both ordinary and phantom global monopoles. In Figure (\ref{EMplots2}), we show the variation of the energy emission rate with frequency $\omega$ for both ordinary and phantom global monopoles with different values of $\alpha=0.1$, $0.01$, $0.001$ considering small values of $\eta=0.1$. We see that the energy emission rate varies for ordinary global monopole compared to phantom global monopole.
    
\subsection{Geodesics Equations}\label{subsec:4}

\begin{figure}[htb!]
\centering
\includegraphics[width=0.35\textwidth,height=0.25\textwidth]{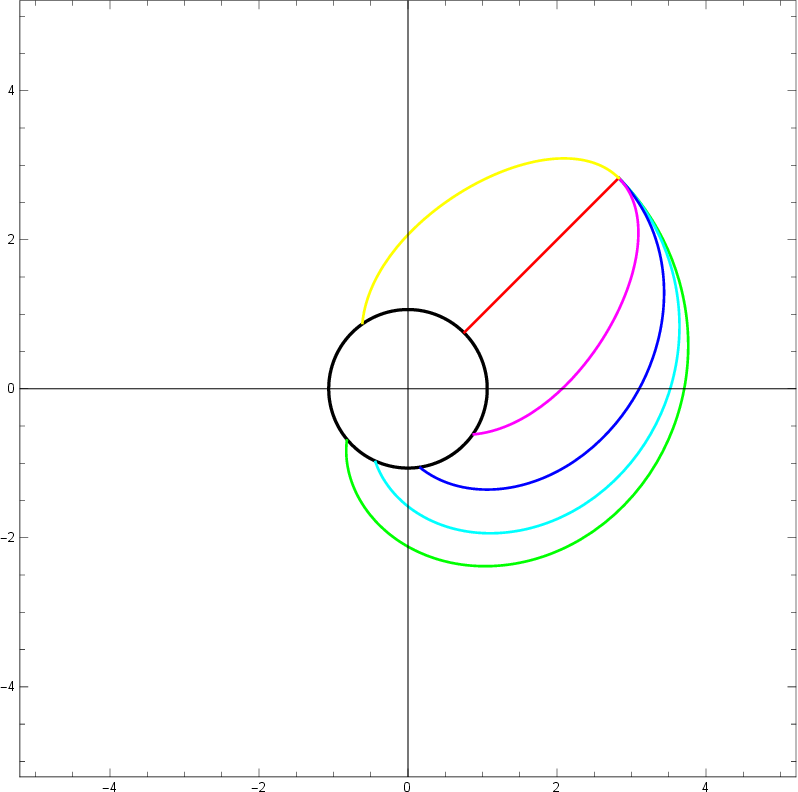}\quad\quad
\includegraphics[width=0.35\textwidth,height=0.25\textwidth]{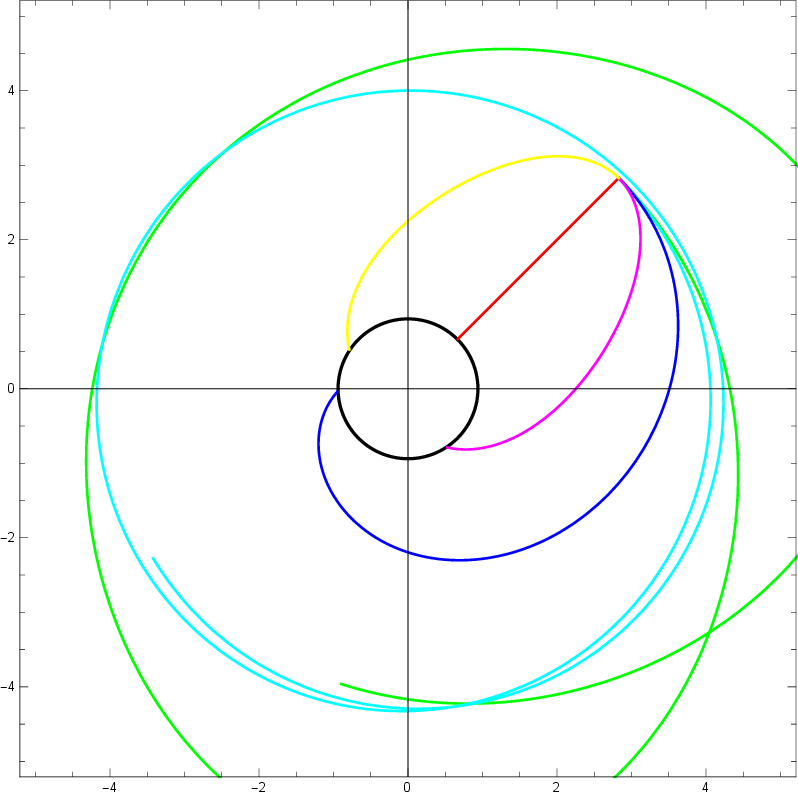}
\hfill\\
\includegraphics[width=0.35\textwidth,height=0.25\textwidth]{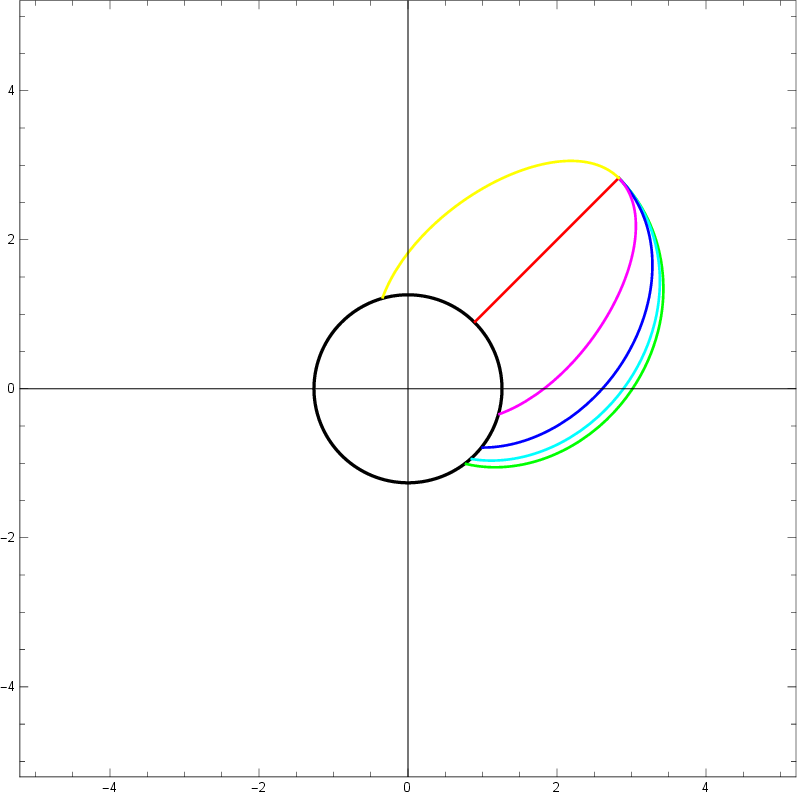}\quad\quad
\includegraphics[width=0.35\textwidth,height=0.25\textwidth]{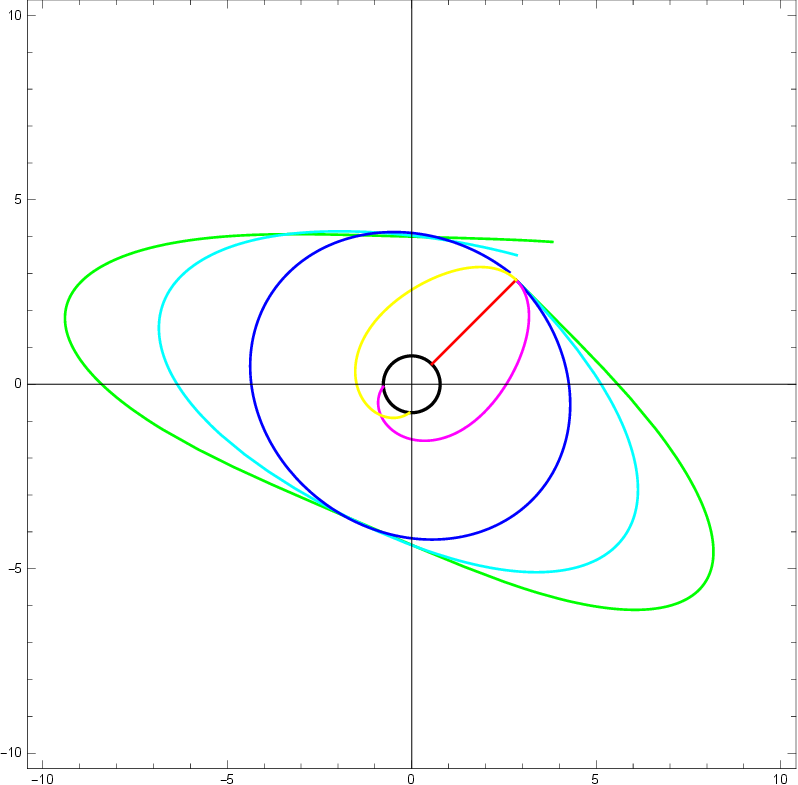}
\caption{ \label{vef} Trajectories of massive particles for slightly different initial
velocities for $M=1$, $P=\frac{3}{8\protect\pi }$, $\protect\alpha =0.01$. Top: in left figure $ \xi=1$ and  $ \eta=0.1$ and in the right one $ \xi=-1$  and  $ \eta=0.1$. Bottom: in the left figure $ \xi=1$  and  $ \eta=0.2$ and the right one $ \xi=-1$  and  $ \eta=0.2$.}
\end{figure}

\begin{figure}[htb!]
\centering
\includegraphics[width=0.35\textwidth,height=0.2\textwidth]{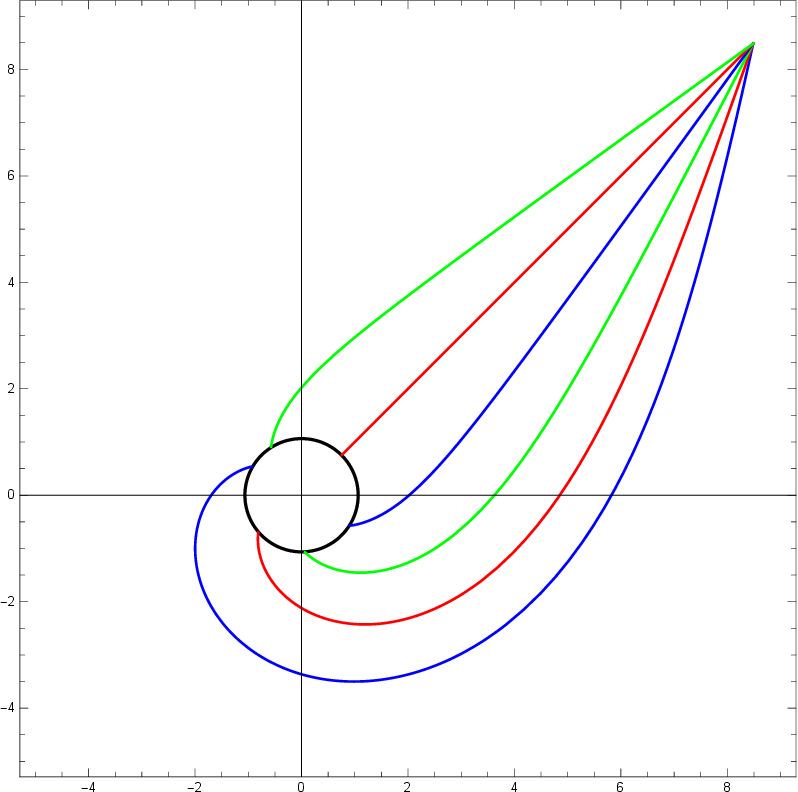}\quad\quad
\includegraphics[width=0.35\textwidth,height=0.2\textwidth]{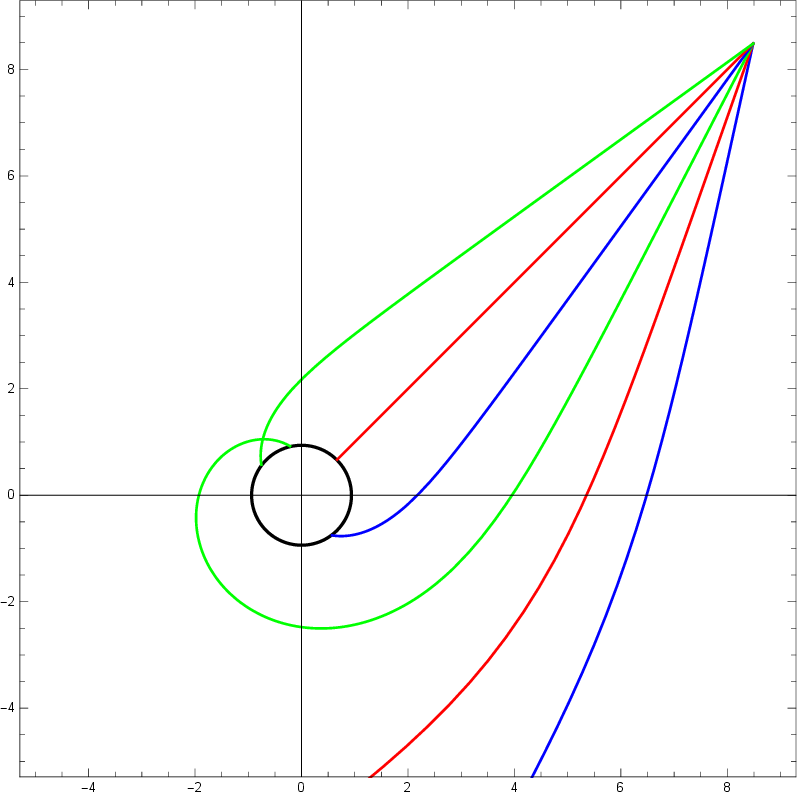}
\hfill\\
\includegraphics[width=0.35\textwidth,height=0.2\textwidth]{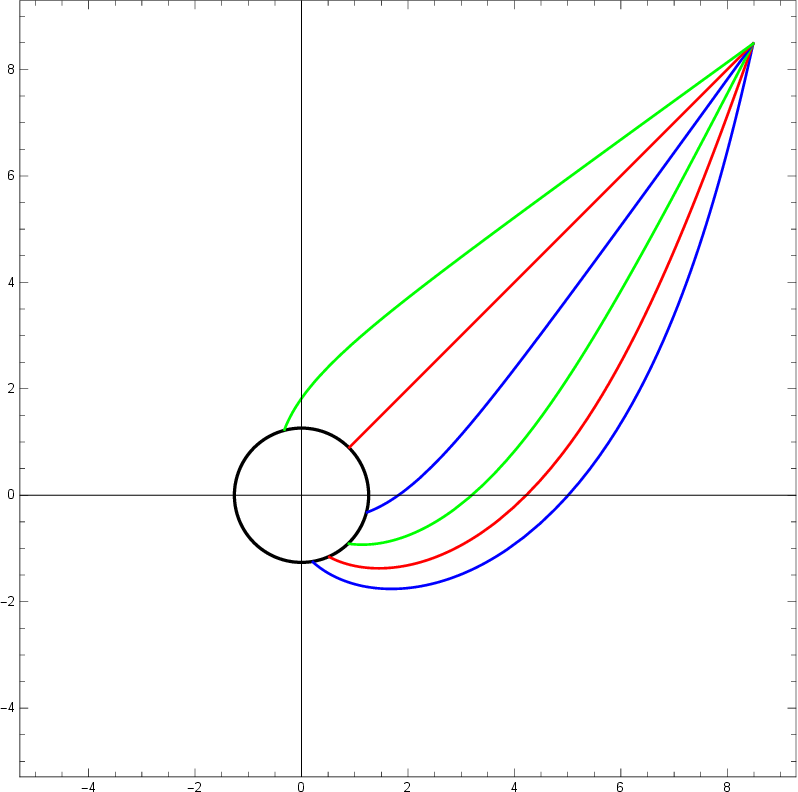}\quad\quad
\includegraphics[width=0.35\textwidth,height=0.2\textwidth]{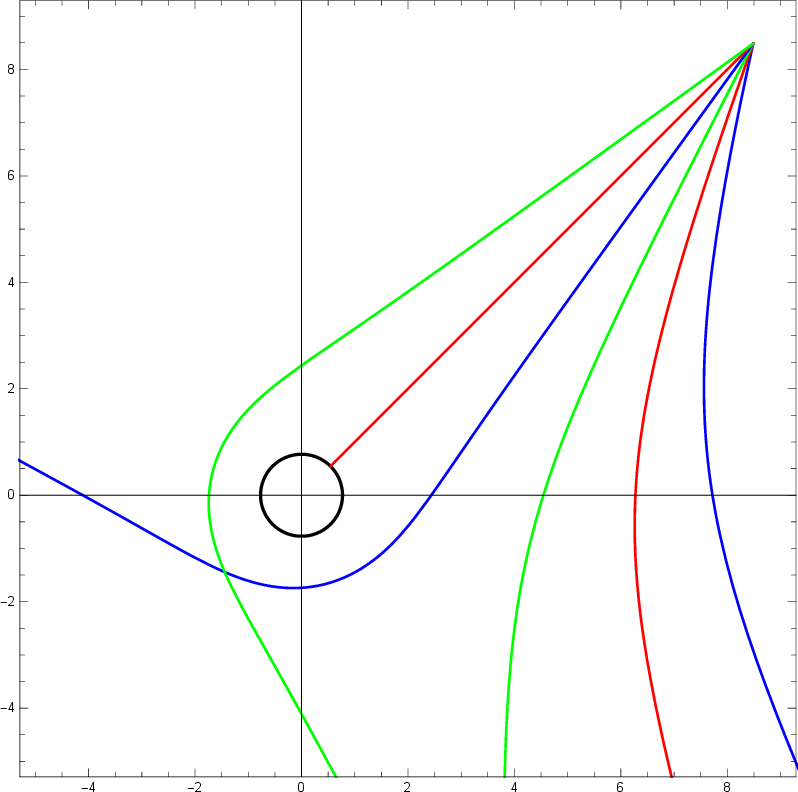}
\caption{\label{geoph} Photon trajectories for slightly different initial
velocities for $M=1$, $r_{0}=0.05$, $P=\frac{3}{8\protect\pi }$, $\protect\alpha =0.01$. Top: in the left figure $\xi=1$ and $\eta=0.1$ and the right one $ \xi=-1$ and $ \eta=0.1$. Bottom: in the left figure $\xi=1$ and $\eta=0.2$ and the right one $ \xi=-1$ and $\eta=0.2$}
\end{figure}

Geodesics hold significant importance in physics, as they unveil the curvature of spacetime and the behavior of particles under gravitational forces. Understanding geodesics in the presence of quantum corrections can provide insights into how particles and fields behave at very small scales, where quantum effects become dominant. Furthermore, it is essential to understand the geodesic structure with quantum corrections present in order to interpret and analyze astrophysical observations associated with black holes, such as the characteristics of accretion disks and shadows. { To examine the null and time-like geodesics of test particles around the vicinity of the BH, we begin by deriving the geodesic equations along with their corresponding constraint equations \cite{GM9,GM10,GM11}}
\begin{equation}
\frac{d^{2}x^{\beta }}{d\tau ^{2}}+\Gamma _{\mu \nu }^{\beta }\frac{dx^{\mu }%
}{d\tau }\frac{dx^{\nu }}{d\tau }=0,
\end{equation}%
\begin{equation}
g_{\mu \nu }\frac{dx^{\mu }}{d\tau }\frac{dx^{\nu }}{d\tau }=\epsilon .
\end{equation}%
Here $\epsilon =0$ and $\epsilon =-1$ correspond to null and time-like geodesics, respectively. For the metric (\ref{b1}), the geodesic equations take the following forms, 
\begin{equation}
t^{\prime \prime }\left( \tau \right) =-\frac{r^{\prime }t^{\prime }\left(
3\left( \alpha ^{2}+2M\sqrt{r^{2}-\alpha ^{2}}\right) +8\pi P\left( -\alpha
^{4}+2r^{4}-\alpha ^{2}r^{2}\right) \right) }{r\left( 8\pi Pr^{4}-3\alpha
^{2}-6M\sqrt{r^{2}-\alpha ^{2}}+8\pi \alpha ^{4}P+r^{2}\left( 3-16\pi \alpha
^{2}P\right) -24\pi \eta ^{2}\xi r\sqrt{r^{2}-\alpha ^{2}}\right) }.
\end{equation}%
\begin{eqnarray}
&&r^{\prime \prime }(\tau )=-\frac{1}{18r^{3}\sqrt{r^{2}-\alpha ^{2}}}\Big[%
t^{\prime 2}\Big\{3(\alpha ^{2}+2M\sqrt{r^{2}-\alpha ^{2}})+8\pi P(-\alpha
^{4}+2r^{4}-\alpha ^{2}r^{2})\Big\}\times   \notag \\
&&\Big\{-6M+8\pi P\left( r^{2}-\alpha ^{2}\right) ^{3/2}+3(\sqrt{%
r^{2}-\alpha ^{2}}-8\pi \eta ^{2}\xi r)\Big\}\Big]  \notag \\
&&+\left( \sin ^{2}\theta \phi ^{\prime 2}-\theta ^{\prime 2}\right) \left(
-2M+\frac{8\pi }{3}P\left( r^{2}-\alpha ^{2}\right) ^{3/2}+\left( \sqrt{%
r^{2}-\alpha ^{2}}-8\pi \eta ^{2}\xi r\right) \right)   \notag \\
&&-\frac{r^{\prime 2}\Big\{-3r \Big(\alpha ^{2}+2M\sqrt{r^{2}-\alpha ^{2}}\Big)  -8\pi P(r^{2}-\alpha ^{2})(2r^{3}+\alpha ^{2}r)\Big\}}{2r^{2}\sqrt{r^{2}-\alpha ^{2}}\Big\{-6M+8\pi P(r^{2}-\alpha
^{2})^{3/2}+3(\sqrt{r^{2}-\alpha ^{2}}-8\pi \eta ^{2}\xi r)\Big\}}.
\end{eqnarray}%
\begin{equation}
\text{$\theta $}^{\prime \prime }\left( \tau \right) =\sin \theta \cos
\theta \phi ^{\prime 2}-\frac{2\theta ^{\prime }r^{\prime }}{r}.
\end{equation}%
\begin{equation}
\text{$\phi $}^{\prime \prime }\left( \tau \right) =-\frac{2\phi ^{\prime
}\left( r^{\prime }+r\theta ^{\prime }\cot \theta \right) }{r}.
\end{equation}%
It's clear that the equations above do not have analytical solutions, thus
necessitating numerical analysis to investigate the trajectories of a test
particle along geodesics. To achieve this, we use Mathematica code
(Kerr Orbit GR Project), and we choose specific field parameters, such as $%
\theta ,$ $\phi $, then we proceed to solve for them. Initially, we provide
initial conditions and solve for the coordinates parameterized with respect
to $\tau $. The Figure \ref{vef} is made with initial coordinates $x^{\mu
}\left( 0\right) =\left\{ 0,4,\frac{\pi }{2},\frac{\pi }{4}\right\} $ and
different initial velocities. Notably, the energy scale of symmetry-breaking 
$\eta $, and kinetic energy $\xi $ parameters have a significant impact on
trajectories of massive particles, causing a contraction\ to it as $\eta $
increases for fixed $\xi $. Furthermore, for $\xi =-1,$ we observe that the
event horizon expands as $\eta $ decreases, while for $\xi =+1$, the event
horizon expands as $\eta $ increases. Figure \ref{geoph} illustrates the
behavior of trajectory photons. As expected, these particles also
experience significant modifications compared to massive particles. 

\section{Conclusions}\label{sec:3}

In this paper, we conducted a thorough investigation of a spherically symmetric BH featuring ordinary and phantom global monopoles within the framework of quantum-corrected proposals. Firstly, we examined the thermodynamic properties of the BH under consideration in sub-section \ref{subsec:1}. We found that the black hole mass increases in the presence of a phantom monopole, as previously observed for non-AdS black holes \cite{SCJJ, AHEP}, with quantum corrections leaving this effect largely unchanged, except for introducing a lower bound on the event horizon mass due to the square root term. We noted that the quantum effects primarily affect the black hole mass on very small event horizons similar to \cite{wu, Biz2023}, while the overall structure and behavior at larger scales remain consistent with classical predictions. Our examination of the Hawking temperature revealed that the monopole term contributes as a shift that either decreases or increases inversely with the event horizon, in agreement with the results of \cite{Xian}. Additionally, we found that due to quantum corrections, the black hole cannot radiate below a critical horizon value, which aligns with the findings presented in \cite{Biz2023}. It is important to emphasize that the effects of quantum corrections on the Hawking temperature are significant only for relatively small BHs \cite{Biz2023}. Similar to the effect on the Hawking temperature, the presence of the global monopole induces a shift in the specific heat function, which corresponds to a second-order phase transition occurring at different event horizons. Following the literature \cite{BH1, Biz2023}, we found that, due to quantum corrections, the BH is unstable between specific event horizon radii that are distinct from zero. We demonstrated that while the entropy function preserves its conventional form, the event horizon is influenced by quantum corrections in addition to the energy scale and kinetic energy terms. We then showed that the presence of the monopole alters the equation of state function of the BH. Using qualitative representations, we demonstrated that this impact does not change the characteristics of the equation of state for relatively larger BHs. On the other hand,  we showed that the type of monopole plays an important role; for instance, in the ordinary global monopole case, the maximal pressure decreases as the quantum correction parameter value increases. In contrast, for the phantom global monopole case, the maximal pressure increases with the $\alpha$ value, and in the absence of a monopole, quantum corrections do not affect the maximal pressure value \cite{BH1}. All relevant quantities, including the mass function, Hawking temperature, heat capacity, and pressure isotherms, are depicted as functions of the horizon radius for various values of the parameter $\alpha$ in the presence of ordinary, phantom, and no phantom global monopoles, as shown in Figures (\ref{Mplots1})–(\ref{Pplots2}).

Moreover, we analyzed the effective potential of the system for both ordinary and phantom global monopoles in sub-section \ref{subsec:2}. We demonstrated that the effective potential for null geodesics differs between ordinary and phantom global monopoles. This discrepancy is also evident in the case of massive time-like geodesics. To visualize the effective potential for null and time-like geodesics, we generated Figures (\ref{Vppotentel})--(\ref{Vnnpotentel}) depicting their behavior for different values of the quantum corrected parameter $\alpha$, while setting the angular momentum values to $L=1,2,3$ for both ordinary and phantom global monopoles. Additionally, Figure (\ref{raplots1}) illustrates the variation of the radius of the BH shadow. In Tables \ref{tab1} and \ref{tab2}, we calculated a few values of the photon radius ($r_{p}$) and impact parameter $(\beta+\delta^2)$ with different values of the quantum corrected parameter $\alpha$, while considering $\eta=0.1, 0.01$ for both ordinary and phantom global monopoles. In addition, we computed the energy emission rate from the BH in sub-section \ref{subsec:3}. To delve deeper into this phenomenon, we depicted the emission rate with frequency $\omega$ in Figures (\ref{EMplots1})--(\ref{EMplots2}) for various values of the quantum correction parameter $\alpha$, considering both ordinary and phantom global monopoles. It becomes evident from these figures that the emission rates differ between these two types of monopoles. Lastly, we investigated the geodesic equations of motion in sub-section \ref{subsec:4}. To visually represent the trajectories of photons and massive particles, we created Figures (\ref{vef})--(\ref{geoph}), illustrating the effects of ordinary and phantom global monopoles for a constant value of the quantum corrected parameter $\alpha$ and the cosmological constant parameter $\ell$.

A compelling junction between quantum gravity and thermodynamics is established by studying the geodesic trajectories and thermodynamic features of test particles around quantum-corrected AdS black holes with phantom global monopoles. The implications of these phenomena have been the focus of recent research, particularly regarding how the behavior of black holes in anti-de Sitter (AdS) space is influenced by the presence of global monopoles. Within an AdS backdrop, spherically symmetric quantum-corrected black hole space-time is described by a novel metric ansatz. This approach allows for a comprehensive examination of the impact of both conventional and phantom global monopoles on the black hole's parameters. AdS black holes are crucial for theoretical physics, providing a framework to study quantum phenomena in curved space-time, especially in the context of gravity duality and the holographic principle. The presence of phantom global monopoles, which are exotic matter configurations with negative energy density, complicates the structure of space-time and may influence the thermodynamic behavior and stability of black holes.

{ Understanding the geodesic trajectories of test particles around these black holes provides valuable insights into the extreme gravitational and space-time properties. It sheds light on the behavior of particles within varying gravitational fields resulting from quantum corrections, potentially revealing new aspects of black hole thermodynamics, including temperature and entropy variations that differ from classical predictions.} Additionally, these studies may have implications for the dynamics of cosmic structures and the evolution of the universe, especially in the context of dark energy and exotic matter. Beyond theoretical considerations, this research is important because it could illuminate the fundamental properties of gravity and astrophysical phenomena. By investigating these complex systems, scientists aim to deepen our understanding of black hole thermodynamics and potentially uncover novel physics that bridges general relativity and quantum mechanics. In doing so, this work contributes to our broader effort to comprehend the fundamental processes that govern the universe, opening new avenues for exploration into the structure of space-time and the ultimate fate of black holes.

In our future work, we will focus on the deflection angle of photon rays in this black hole (BH) model, investigating the influence of various factors such as global monopoles, quantum corrections, and the cosmological constant associated with the parameter $\ell$. Despite the considerable advances in black hole physics since the groundbreaking discoveries of black hole thermodynamics and Hawking radiation, the quest for a unified theory of quantum gravity remains one of the most formidable challenges in theoretical physics. { The exact derivation of Hawking effects in dynamical space-times continues to evade us, as does the generalization of thermodynamic laws in such contexts, based on first principles. Numerous definitions of horizons and surface gravity have been proposed to extend black hole mechanics to dynamic scenarios, yet the debate persists over which horizon truly exhibits thermodynamic properties and which form of surface gravity is directly linked to Hawking temperature. Moreover, each Hawking particle emitted from a black hole induces a back-reaction effect on the source, continuously modifying the space-time geometry. Addressing these complexities, particularly the interplay between back-reaction and dynamic space-time evolution will be a key focus of our future endeavors.}

\section*{Conflict of Interests} 

The authors declare no such conflict of interest.

\section*{Data Availability Statement} 

No data were generated or analyzed in this study.

\section*{Acknowledgements}

F.A. acknowledges the Inter-University Centre for Astronomy and Astrophysics (IUCAA), Pune, India for granting visiting associateship. B. C. L. is grateful to Excellence Project PřF UHK 2211/2023-2024 for the financial support.

\end{document}